\documentclass[aps,prd,superscriptaddress,preprintnumbers,floatfix,nofootinbib,notitlepage,showkeys,showpacs, twocolumn]{revtex4-1}

\usepackage[utf8]{inputenc}

\usepackage{graphicx}
\usepackage{hyperref}
\usepackage{latexsym}
\usepackage{amsmath}
\usepackage{amssymb}
\usepackage{bbm}
\usepackage{ulem}
\usepackage{pdfsync}
\usepackage{epsfig}
\usepackage{epstopdf}
\usepackage{subfigure}
\usepackage{xcolor}
\usepackage{comment}
\usepackage{slashed}
\usepackage{multirow}
\usepackage{xfrac}
\usepackage{physics}
\usepackage{cleveref} 

\newcommand{\avg}[1]{\left\langle #1 \right\rangle}

\newcommand{\GSP}[1]{\textcolor{orange}{[{\bf GSP}: #1]}}
\newcommand{\MZ}[1]{\textcolor{red}{[{\bf MZ}: #1]}}
\newcommand{\new}[1]{\textcolor{blue}{#1}}
\begin{document}

\title{Summary statistic for pulsar timing arrays}

\author{Gabriela Sato-Polito}
\email{gsatopolito@ias.edu}
\affiliation{School of Natural Sciences, Institute for Advanced Study, Princeton, NJ 08540, United States}

\author{Matias Zaldarriaga}
\affiliation{School of Natural Sciences, Institute for Advanced Study, Princeton, NJ 08540, United States}

\author{Barak Zackay}
\affiliation{Department of Particle Physics \& Astrophysics, Weizmann Institute of Science, Rehovot 76100, Israel}

\begin{abstract}
The timing residuals produced by gravitational-wave signals can be described as an incoherent (pulsar term) contribution and a coherent (Earth term) map on the sky, which PTAs measure at the locations of the timed pulsars. The observed Earth term map and the variance induced by the pulsar term contain all of the information about any GW signal available to a PTA (assuming pulsar distances are unknown). Furthermore, any type of signal produces on average the same angular correlation function, the Hellings and Downs curve, which decays steeply with multipole as $C_\ell \propto 1/[(\ell+2)(\ell+1)\ell(\ell-1)]$. This suggests that the signal is inherently low-dimensional and therefore only a small number of parameters are needed to fully characterize it. We present an expression for the PTA likelihood that makes the dependence on the Earth term map and pulsar term variance explicit, and show that only a few spherical harmonic coefficients are needed to capture most of the information about the signal. To quantify this in a realistic setting, we compute the Fisher matrix of the amplitude of a stochastic background or a deterministic point source assuming the noise properties and sky locations of the pulsars in the NANOGrav 15yr dataset. We find that $\ell_{\rm max}=2$ of the Earth term map and the monopole of the pulsar term variance retain $\sim 95\%$ of the information about the signal. For a point source, including the dipole of the pulsar term variance is important to achieve a similar fraction.
\end{abstract}

\maketitle
\section{Introduction}
Multiple pulsar timing array (PTA) collaborations reported evidence for low-frequency gravitational waves (GWs) \cite{NANOGrav_stoc, EPTA:2023fyk, PPTA, CPTA, Miles:2024rjc} for the first time, with subsequent data releases and combined data sets expected to achieve a $5\sigma$ detection threshold in the near future. A GW signal produces a shift in the arrival times of pulses that depends on the metric perturbation on Earth and at each pulsar. The former produces a timing residual that is coherent across pulsars, the Earth term, and whose angular correlation is the Hellings and Downs (HD) curve~\cite{HD}, leading to the characteristic quadrupole-dominated signature of the GW signal.

While the origin of the observed signal is still unknown, the sources are expected to be inspiralling supermassive black hole binaries (SMBHBs). Following a significant detection, a major observational goal is to resolve the GW sky and potentially detect individual sources \cite{Sesana:2008xk,Rosado:2015epa,Kelley:2017vox}. Since on average any GW signal produces the same Hellings and Downs correlation function, doing so requires new data analysis tools beyond those in use for the isotropic stochastic background searches~\cite{Sesana:2010ac, Corbin:2010kt, 2012ApJ...756..175E, 2012PhRvD..85d4034B, Becsy:2022zbu,NANOGrav_individual, NANOGrav_anisotropy, EPTA_continuous}. These new tools are typically categorized in the literature as template-based or anisotropy searches \cite{Cornish:2013nma}, and have been used to characterize the angular structure of the PTA signal~\cite{Mingarelli:2013dsa, Taylor:2013esa, Cornish:2014rva, Gair:2014rwa, Hotinli:2019tpc, Ali-Haimoud:2020iyz, Ali-Haimoud:2020ozu}. 

We can think of the Earth and pulsar terms as maps on the sky, which PTAs measure at the locations of the timed pulsars. These observed maps contain all of the available information about the GW signal. Regardless of the nature of the source, the expectation value of the power spectrum of the Earth term map is identical and strongly dominated by the quadrupole term, decaying as a function of multipole as $C_\ell \propto 1/\left[(\ell+2)(\ell+1)\ell(\ell-1)\right]$. The fast decay suggests that in general one only needs to measure a small number of parameters to fully characterize the signal. 

In this work, we explore the use of the spherical harmonic coefficients of the Earth and the pulsar term variance maps as a nearly lossless summary statistic for PTAs. We begin by deriving a new expression for the PTA likelihood, in which our ignorance of distances to pulsars is translated into a Gaussian prior on the timing residual due to the pulsar term and we marginalize over it. The Earth term map and the variance of the pulsar term are then expanded in spherical harmonics, which comprise the proposed summary statistic. While the harmonic expansion of GW signals in PTAs has been explored in the literature~\cite{Gair:2014rwa, Roebber:2016jzl, PhysRevD.110.043043,Nay:2023pwu, NANOGrav_harmonic}, we focus on their potential to compress the data set while retaining most of the information regarding GW signals.

We quantify how well a small number of spherical harmonic coefficients can capture the information about the signal by computing the Fisher matrix for the amplitude of a gravitational-wave background (GWB) or a deterministic point source, and compare full and compressed data sets. We consider both an idealized toy-model and a realistic estimate based on noise properties and sky locations of the pulsars in the NANOGrav 15yr (NG15) dataset~\cite{NANOGrav_15yr_dataset}. From this exercise, we conclude that a measurement with $\ell_{\rm max}=2$ of the Earth term map and the monopole of the pulsar term variance in the first few frequency bins of a PTA contains most ($\sim 95\%$) of the information about the signal. For a point source, we show that the dipole in the pulsar term variance contributes significantly to the Fisher information. While the fraction of the signal-to-noise retained in the truncation depends on the source sky location, we show that an $\ell_{\rm max} = 3$ and the pulsar term up to the dipole are sufficient to achieve $\geq 90\%$ of the information about the signal across the entire sky.

This paper is organized as follows. In Sec.~\ref{sec:likelihood} we derive a modified PTA likelihood in which the Earth-term map and the pulsar-term variance appear explicitly, with the Gaussian approximation for the pulsar-term prior examined in detail in Appendix~\ref{app:gauss_approx}. In Sec.~\ref{sec:EP_SH} we present the expected Earth- and pulsar-term spherical harmonic coefficients for a single source and for a stochastic background. In Sec.~\ref{sec:FIM} we assess their detectability, deriving the Fisher matrix of the maps and of the signal amplitude for both an idealized toy-PTA and a realistic estimate based on the noise properties and sky locations of the NANOGrav 15yr pulsars. Finally, in Sec.~\ref{sec:harm} we quantify how much of the information about a stochastic background or a deterministic point source is retained when the maps are truncated to the first few multipoles.

\section{PTA Likelihood}\label{sec:likelihood}
We begin by considering a simple scenario in which the timing residual data for a single pulsar $i$ at a single frequency $f$ measured by PTAs can be described as the sum of the residual produced by gravitational waves $z_i = z_{i,e} + z_{i,p}$ and noise $n_i$
\begin{equation}
d_i = z_{i,e} + z_{i,p} + n_i,
\label{eq:d_i}
\end{equation}
where subscripts $e$ and $p$ denote the Earth and pulsar terms of the timing residuals and $n_i$ is a Gaussian noise contribution with mean zero and variance $\sigma^2_{i,n}$. Under this assumption, we can always write the likelihood as
\begin{equation}
p(d_i|z_{i}, \sigma^2_{i,n}) = \frac{1}{2\pi\sigma^2_{i,n}} \exp\left\{-\frac{|d_i-z_{i,e}- z_{i,p}|^2}{2 \sigma^2_{i,n}}\right\}.
\end{equation}
Note that here we assume that $d$ is a complex number, where each component has variance $\sigma_{i,n}^2$, hence the normalization above. Since the noise is assumed to be independent, the likelihood for multiple pulsars is just $\prod_i p(d_i|z_{i}, \sigma^2_{i,n})$.

For a single GW source, which we label $s$, the pulsar term differs from the Earth term by a phase $\psi_{i,s}$
\begin{equation}
z^s_{i,p} = z^s_{i,e} e^{i\psi_{i,s}},
\end{equation}
which is given by
\begin{equation}
\psi_{i,s} = \frac{2\pi f L_i}{c}(1+\hat{\Omega}_s\cdot \hat{n}_i),
\end{equation}
where $L_i$ is the distance to the pulsar, $\hat{\Omega}_s$ is the angular position of the source, and $\hat{n}_i$ is the angular position of the pulsar. In the presence of multiple sources, the timing residual is given by the sum of the contribution of each source
\begin{equation}
z_{i,e} = \sum_s z_{i,e}^s, \quad \text{and} \quad
z_{i,p} = \sum_s z_{i,e}^s e^{i\psi_{i,s}}.
\end{equation}

In the regime where the distance to the pulsar is not known to a precision comparable to the wavelength of the GW, the prior on $\psi_{i,s}$ is a uniform distribution. This is an accurate characterization of the distance measurement errors for almost all pulsars currently included in PTA datasets~\cite{NANOGrav_individual, Charisi:2023rdl}. We will approximate the pulsar term $z_{i,p}$ as a Gaussian random variable
\begin{equation}
    p(z_{i,p}|\sigma^2_{i,p}) = \frac{1}{2\pi \sigma^2_{i,p}} e^{-|z_{i,p}|^2/2\sigma^2_{i,p}}
    \label{eq:zp_gauss}
\end{equation}
where the variance is defined per real/imaginary component, such that
\begin{equation}
\avg{z^*_{i,p} z_{i,p}}_{\psi_{i,s}} = \sum_s |z^s_{i,e}|^2 \equiv \ 2\sigma^2_{i,p}.
\label{eq:zp_variance}
\end{equation}
where the average is over the pulsar term phase $\psi_{i,s}$. Marginalizing over the pulsar term leads to the likelihood
\begin{equation}
\begin{split}
p(d_i|z_{i,e}, \sigma^2_{i,p}) =& \int dz_{i,p} \, p(d_i|z_{i,e}, z_{i,p}) \, p(z_{i,p}|\sigma^2_{i,p}) \\
=& \frac{1}{2\pi \sigma^2_i} \exp\left\{-\frac{1}{2}\frac{|d_i-z_{i,e}|^2}{\sigma^2_i}\right\},
\label{eq:p_gauss}
\end{split}
\end{equation}
where $\sigma^2_i \equiv \sigma_{i,n}^2 + \sigma^2_{i,p}$. The Earth and pulsar terms, $z_{e}(\hat{n})$ and $\sigma^2_{p}(\hat{n})$, can be thought of as maps on the sky, which we observe at the pulsar locations $\hat{n}_i$. It will be convenient to express the two maps using their spherical harmonic coefficients, $a_{\ell m}$ and $b_{\ell m}$, respectively, for the full PTA:
\begin{equation}
\begin{split}
 \log\big[p(\mathbf{d}|\mathbf{a},\mathbf{b}, \sigma^2_h)\big] =& -\sum_i \log(2\pi \sigma^2_i ) \\ &-\frac{1}{2} \sum_i \frac{|d_i -  \sigma_h \sum_\alpha a_\alpha Y_{\alpha,i}|^2}{\sigma^2_i},
 \label{eq:loglike}
\end{split}
\end{equation}
and 
\begin{equation}
    \sigma^2_i = \sigma^2_h \sum_\beta b_\beta Y_{\beta,i}  + \sigma^2_{n,i},
\end{equation}
where we use the monopole of the pulsar term $\sigma^2_h$ as a scale in order to work in dimensionless variables. That is, we factorize
\begin{equation}
    \sigma^2_h \equiv \int \frac{d^2\hat{n}}{4\pi} \sigma^2_p(\hat{n}),
    \label{eq:sig2h_def}
\end{equation}
and $\alpha$, $\beta$ correspond to $(\ell m)$ that are summed over. The coefficients $a_{\ell m}$ and $b_{\ell m}$ efficiently summarize the PTA dataset into a small number of parameters.

The main step in the preceding discussion that requires attention is the approximation in Eq.~\ref{eq:zp_gauss}. While it is evidently satisfied in the limit of many sources thanks to the central limit theorem, its validity is less clear for few or a single dominant source. There are three effects that are relevant to establish the validity of the Gaussian assumption: the noise contribution $n_i$, the contribution from many sources that produce a Gaussian contribution that can be modelled as an isotropic stochastic background, and the fact that estimating $a_{\ell m}$ and $b_{\ell m}$ sums over multiple pulsars. At low frequencies, a larger effective number of sources are expected to dominate the signal, resulting in a Gaussian contribution from the stochastic background. At higher frequencies, where a single source might dominate the signal, one has larger noise contributions and more pulsars measured at those frequencies. 

In Appendix~\ref{app:gauss_approx}, we investigate the accuracy of the approximation given in Eq.~\ref{eq:zp_gauss}. We do so by focusing on the extreme case in which only a single source dominates the signal (ignoring a stochastic background contribution from many fainter sources), and consider both a toy-PTA in which all pulsars have identical noise properties and are uniformly distributed on the sky and a realistic estimate based on the NANOGrav 15yr pulsars at the most sensitive frequency bin. The realistic noise estimates are discussed in more detail in Sec.~\ref{sec:sens}. We show that the Gaussian assumption is an excellent approximation for current PTAs.


\section{Earth and pulsar term spherical harmonic coefficients}\label{sec:EP_SH}
\begin{figure}[t]
    \centering
    \includegraphics[width=0.95\linewidth]{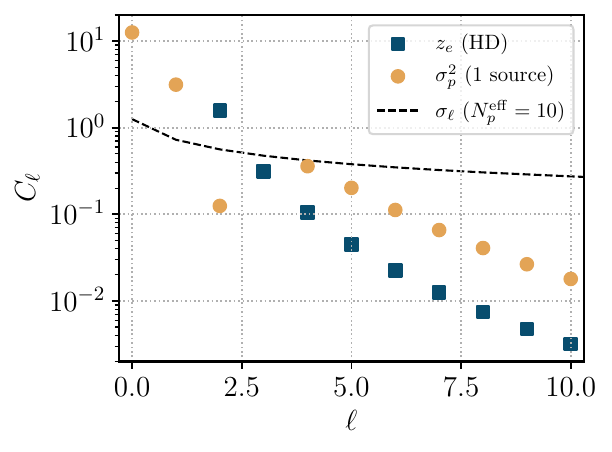}
    \caption{Angular power spectra of the Earth-term map $z_e(\hat{n})$ (blue squares) and the pulsar-term variance map $\sigma^2_p(\hat{n})$ (yellow dots) in the dimensionless normalization of Eqs.~\ref{eq:cl_ze} and \ref{eq:bLM}. The Earth-term spectrum (the harmonic-space counterpart of the HD correlation) is identical for a single source and for an isotropic stochastic background, while the pulsar-term variance map is not. The case shown in yellow correponds to a single edge-on source. The dashed line shows the noise per multipole $\sigma_{\ell} = N_\ell/\sqrt{2\ell+1}$ (defined in Eq.~\ref{eq:lambda_ell}) for an effective number of pulsars $N^{\rm eff}_p = 10$.}
    \label{fig:cl}
\end{figure}

We begin by expanding the Earth term map $z_e$ in spherical harmonics
\begin{equation}
z_e(\theta,\phi) = \sigma_h \sum_{\ell m} a_{\ell m} Y_{\ell m}(\theta,\phi).
\end{equation}
Note that $z_e$ is a complex field on the sphere and therefore each $\ell$ is described by $2(2\ell + 1)$ numbers. Following the notation of Ref.~\cite{Roebber:2016jzl}, we first consider the case where the signal is produced by a single source. If the source is located on the $\hat z$ axis, the Earth term map it produces can be written in angular coordinates as
\begin{equation}
z_{e}(\theta,\phi) = \frac{1}{4} (1+\cos \theta) \left[h_2 e^{2i\phi} + h_{-2} e^{-2i\phi}\right],
\label{eq:z1e}
\end{equation}
where we have defined $h_{\pm 2} = h_+ \pm i h_\times$. For a circular binary, the coefficients $h_{\pm 2}$ are related to the physical parameters of the binary by
\begin{equation}
h_{\pm 2} (f) = \frac{{\cal A} (a\pm b)}{2} e^{\pm 2 i \psi} e^{i\Phi_0} \delta_D(f-f_s),
\end{equation}
where $f$ are positive frequencies, $\psi$ is the polarization angle, $\Phi_0$ is the phase, and the parameters ${\cal A}$, $a$, and $b$ are defined in terms of the chirp mass $\mathcal{M}_c$, the luminosity distance $d_L$, the rest-frame frequency $f_r$, and the inclination angle of the orbital plane with respect to the line of sight $\iota$, as
\begin{equation}
\begin{split}
    {\cal A}=& 2\frac{(G\mathcal{M}_c)^{5/3}}{c^4 d_L} (\pi f_r)^{2/3}, \text{and} \\ a=&1+\cos^2\iota, \quad b=-2\cos\iota.
\end{split}
\end{equation}
The spherical harmonics transform of $z_{e}$ results in the coefficients
\begin{equation}
a_{\ell m} = 2\pi \sqrt{\frac{2l+1}{4\pi}\frac{(l-2)!}{(l+2)!}} \frac{h_{\pm 2}}{\sigma_h} \equiv \frac{z_{\ell}}{2} \frac{h_{\pm 2}}{\sigma_h}, \quad \text{for} \quad m=\pm 2,
\label{eq:alm_1source}
\end{equation}
where the last equality defines $z_\ell \equiv \sqrt{4\pi (2\ell+1) \frac{(\ell-2)!}{(\ell+2)!}}$~\footnote{The definition of $z_\ell$ is different from that of Ref.~\cite{Hotinli:2019tpc} by a factor of $(-1)^{\ell}$ because we follow the notation of Ref.~\cite{Roebber:2016jzl}, which defines the GW propagation vector opposite to \cite{Hotinli:2019tpc}.}. In the absence of frequency evolution, an individual source is therefore fully characterized by 6 parameters, corresponding to the 2 complex numbers $h_{\pm 2}$ and the sky location $\theta, \phi$. For a single source Eqs.~\ref{eq:sig2h_def} and \ref{eq:z1e} imply that $\sigma^2_h = \left(|h_{2}|^2 + |h_{-2}|^2\right)/24$. 

The power spectrum of the Earth term map of an individual source or the expectation value of a stochastic background is the Hellings and Downs curve, which in harmonic space is given by \cite{Roebber:2016jzl, Hotinli:2019tpc}
\begin{equation}
    \begin{split}
    C_{\ell} =& \frac{1}{2(2\ell +1)}\sum_m |a_{\ell m}|^2 = \frac{3 z^2_{\ell}}{2\ell+1} \\ \propto& \frac{1}{\left[(\ell+2)(\ell+1)\ell(\ell-1)\right]},
    \label{eq:cl_ze}
    \end{split}
\end{equation}
Note that $C_\ell$ is the variance per real/imaginary component of $a_{\ell m}$. This standard result for the harmonic-space representation of the HD curve is shown as the dark blue squares in Fig.~\ref{fig:cl}.  For the harmonic space description of the Earth term, we can think of the distinction between individual sources and stochastic backgrounds as follows. In the case where there is a dominant source, there is a rotation of the sky coordinates such that all of the power is contained in the $m=\pm 2$ modes and this orientation is aligned across different multipoles $\ell$, while in the stochastic case, each $m$ is identically distributed and drawn from a Gaussian.

We can also think of the pulsar term variance $\sigma^2_p(\theta,\phi)$ as a map. For an individual source, the pulsar term map is just the square of the Earth term, that is
\begin{equation}
    \begin{split}
    |z_e|^2(\theta,\phi) =&\ \sigma^2_h \sum_{\ell m}\sum_{\ell' m'} a_{\ell m} a_{\ell' m'}^* Y_{\ell m}(\theta,\phi)Y_{\ell' m'}^*(\theta,\phi) \\ =&\ 2 \sigma^2_h \sum_{L M} b_{L M} Y_{L M}(\theta,\phi),
\end{split}
\end{equation}
where $b_{LM}$ are the expansion coefficients of $|z_e|^2$. These can be written in terms of the coefficients of $z_e$ as follows
\begin{widetext}
\begin{equation}
    b_{LM}= \frac{1}{2}(-1)^{-m_2-M} \sum_{\alpha_1}\sum_{\alpha_2} a_{\alpha_1}a_{\alpha_2}^* \sqrt{\frac{(2 l_1 +1) (2 l_2 +1) (2 L +1)}{4\pi}}
    \begin{pmatrix}
    l_1 & l_2 & L \\
    0 & 0 & 0
    \end{pmatrix}
    \begin{pmatrix}
    l_1 & l_2 & L \\
    m_1 & - m_2 & -M
    \end{pmatrix}
\label{eq:aLM_3j}
\end{equation}
\end{widetext}
Or, equivalently, by squaring Eq.~\ref{eq:z1e}
\begin{equation}
    \begin{split}
    |z_{e}|^2(\theta,\phi) =& \left(\frac{1+\cos \theta}{4}\right)^2 \times \\ &\left\{|h_2|^2 + |h_{-2}|^2 + h_2 h_{-2}^* e^{4i\phi} + h_2^* h_{-2} e^{-4i\phi}\right\}.
    \end{split}
\end{equation}
Note that only $M=0,\pm 4$ contribute, and these selection rules are also captured in Eq.~\ref{eq:aLM_3j} since $m_1,m_2 = \pm2$. The $M=0$ piece comes from the first two terms and only contributes to $L=0,1$ and $2$. The $M=4$ piece contributes to $L \geq 4$, and there is no contribution for $L=3$. For $L \leq 2$, the coefficients are given by
\begin{equation}
    \begin{split}
    b_{00} &=  \sqrt{4\pi} \\ 
    b_{10} &=  \frac{\sqrt{3} }{2}  b_{00} \\
    b_{20} &=  \frac{\sqrt{5}}{10}  b_{00}.
    \label{eq:bLM}
    \end{split}
\end{equation}
While for $L \geq 4$, we have that
\begin{equation}
    b_{L,4} =  G_L\, \frac{h_2 h_{-2}^*}{2\sigma^2_h}, \quad \text{and} \quad b_{L,-4} = G_L\, \frac{h_2^* h_{-2}}{2\sigma^2_h},
\end{equation}
with $G_L$ defined as
\begin{equation}
    \begin{split}
    G_L =& \frac{2\pi}{16}\sqrt{\frac{2L+1}{4\pi}\frac{(L-4)!}{(L+4)!}} \int_{-1}^{1} P_L^4 (1+x)^2 dx \\ =& \frac{2\pi}{16}\sqrt{\frac{2L+1}{4\pi}\frac{(L-4)!}{(L+4)!}} (16 L(L+1)- 192).
    \end{split}
\end{equation}
The integral given in the second equality can be computed by using the fact that
\[
P_L^4(x) = (1-x^2)^2 \frac{d^4 }{dx^4}P_L(x),
\]
and integrating by parts. The amplitudes are bound by $|b_{L,\pm4}| \leq 6 G_L$, with the maximum attained for an edge-on binary ($|h_2| = |h_{-2}|$). For a face-on (circularly polarized) source the $M=\pm4$ coefficients vanish and the pulsar-term variance map is exactly band-limited to $L\leq2$. The associated power spectra for the pulsar term in the presence of an individual source are shown as the yellow dots in Fig.~\ref{fig:cl}. Explicitly, we have that
\begin{align}
    C_{1} &=  \frac{C_{0}}{4} \\
    C_{2} &= \frac{C_{0}}{100},
\end{align}
which explicitly shows a relatively large dipole contribution to the pulsar term.

More generally, the pulsar term variance depends on the realization of sources on the sky, as $2\sigma^2_p(\theta,\phi) = \sum_s |z^s_{e}|^2$. If $z^s_{e}$ is modelled as a random variable drawn from an underlying source population model, it follows that $\sigma^2_p$ is also a random variable obtained from the same population model. The coefficients $b_{LM}$ are therefore random variables, but only $L=0$, $M=0$ have non-zero mean once averaged over the source population, since we are assuming that the sources are uncorrelated with each other. That is,
\begin{equation}
\langle b_{LM} \rangle =  \sqrt{4\pi} \delta_{L0} \delta_{M0}.
\end{equation}
The isotropic stochastic background is obtained in the limit of many faint sources, in which case $b_{LM} \rightarrow \langle b_{LM} \rangle$, and corresponds to the standard pulsar term contribution to the timing residual variance. At the level of the pulsar term, we note that the dominant distinction between the individual source and an isotropic stochastic background is the presence of a large dipole in the direction of the source.

\begin{figure*}[t]
    \centering
    \includegraphics[width=0.65\linewidth]{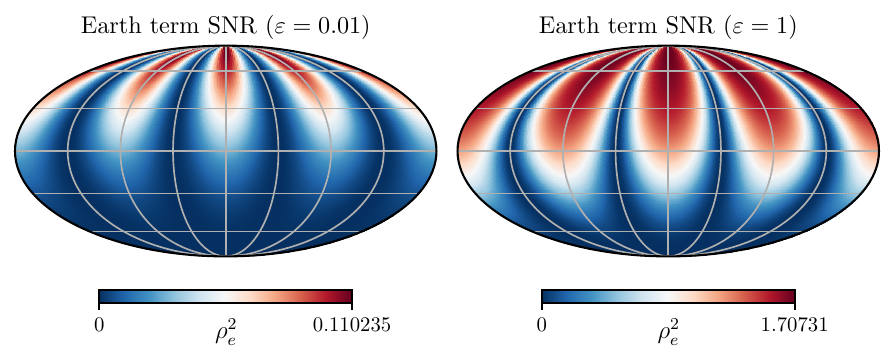}
    \caption{Contribution per pulsar to the signal-to-noise ratio of the Earth term map for $\varepsilon=0.01$ on the left and $\varepsilon=1$ on the right. On the left panel, the SNR map essentially follows the square of the Earth term map, since the pulsar term is negligible compared to the noise, and is dominated by a narrow area around the source location. On the right panel, the pulsar term is comparable to the noise, and the SNR map is more uniform across the sky.}
    \label{fig:SNR_EP}
\end{figure*}

As a final observation, we note that in the single source scenario, the $a_{\ell m}$ and $b_{\ell m}$ coefficients of the Earth and pulsar terms computed above may not necessarily be at the same frequency. For a pulsar with a light-travel time of $\tau$ from Earth and a SMBHB with intrinsic chirp mass $\mathcal{M}_c$, observed frequency $f$, and at a redshift $z$, the shift in the GW frequency between the Earth and the pulsar is
\begin{equation}
\begin{split}
    \Delta f =& \frac{df}{dt} \tau \\ =& 3\ \text{nHz} \left(\frac{(1+z) \mathcal{M}_c}{4\times 10^{9}M_{\odot}}\right)^{5/3} \left(\frac{f}{10\ \text{nHz}}\right)^{11/3} \left(\frac{L_i}{1\text{kpc}}\right),
\end{split}
\end{equation}
where $3$nHz corresponds to the frequency resolution of a PTA with $T_{\rm obs} = 10$yr, the value that the frequency is normalized to is approximately the highest frequency that Ref.~\cite{NANOGrav_stoc} finds evidence for GWs, and the chirp mass normalization corresponds to a total mass $M_{\rm tot} = 10^{10} M_{\odot}$ for an equal mass ratio binary, which is close to the highest masses for SMBH remnants found in the local universe. The approximation that the frequency does not evolve between the Earth and pulsar term is therefore appropriate for current observations, but may fail as PTAs improve sensitivity at higher frequencies.

\section{Sensitivity}\label{sec:FIM}
\subsection{Map Fisher matrix}
The uncertainty on the timing residual maps for a given PTA can be estimated via the Fisher matrix for $a_{\ell m}$ and $b_{\ell m}$,
\begin{align}
    F_{\alpha \beta} \equiv& -\avg{\frac{\partial \ln p}{\partial a_\alpha \partial a_\beta}} = \sum _i w_i\, Y^*_{\alpha, i} Y_{\beta, i}, \label{eq:fisher_aa}\\
    F^p_{\alpha \beta}\equiv& -\avg{\frac{\partial \ln p}{\partial b_\alpha \partial b_\beta}}  = \sum _i w_i^2\, Y^*_{\alpha, i} Y_{\beta, i}, \label{eq:fisher_bb}\\
    F^{ep}_{\alpha \beta} =& -\avg{\frac{\partial \ln p}{\partial a_\alpha \partial b_\beta}} = 0, \label{eq:fisher_ab}
\end{align}
where we defined the pulsar weights $w_i \equiv \sigma^2_h/\sigma^2_i$. It will also be useful to define the ratio between the GW and noise variances as
\begin{equation}
    \varepsilon_i \equiv \frac{\sigma^2_{h}}{\sigma^2_{i,n}}.
    \label{eq:varepsilon}
\end{equation} 
Note that in the case of a stochastic background $\sigma^2_{i,p}/\sigma^2_h = 1$, but not when a single source dominates. The Fisher matrices defined above are the coupling matrix~\cite{Efstathiou:2003dj} between the harmonic coefficients induced by the non-uniform sky distribution and unequal noise. Their inverse corresponds to the noise power spectra $N_{\alpha\beta} = F^{-1}_{\alpha\beta}$ and $N^p_{\alpha\beta} = (F^p)^{-1}_{\alpha\beta}$.

The uncertainties on the maps propagates directly into the sensitivity to GW signals. It is useful to first whiten the data, i.e. rescale it so that the noise covariance becomes the identity, which corresponds to the linear transformation $\mathbf{d} \rightarrow W^{1/2}\mathbf{d}$, where $W_{ij}=\delta_{ij}w_i$ is the matrix of normalized inverse-variance weights. For a signal covariance matrix $S$, the covariance in the whitened basis is given by
\begin{equation}
    M \equiv W^{1/2}\, S\, W^{1/2}, \qquad
    S \equiv \frac{1}{\sigma_h^2}\, \avg{\mathbf{z}_e \mathbf{z}_e^{\dagger}}.
    \label{eq:snr_matrix}
\end{equation}
The eigenvalues of $M$ equal the signal-to-noise ratio of each mode that can be independently determined by the data. The total signal-to-noise ratio of the Earth term map can be written as the trace of $M$,
\begin{equation}
    \rho_e^2= \sum_i \frac{|z_{e,i}|^2}{\sigma_i^2} = \Tr M = \sum_a \lambda_a.
    \label{eq:rho2_e}
\end{equation}
For a point source, we may define the second moment in Eq.~\ref{eq:snr_matrix} as the average over the phase, polarizations, and inclination angles. For a deterministic point source where all parameters are fixed, the average has no effect. Nevertheless, $M$ has a single non-zero eigenvalue equal to the matched-filter SNR$^2$, $\lambda_1 = \rho_e^2$. 

For an isotropic stochastic signal, $S=\Gamma^e$, which is the Hellings and Downs correlation, and in this case excludes the pulsar term contribution to the auto-correlation. In the regime of identical and uniformly distributed pulsars, the Fisher matrix diagonalizes and the eigenmodes of $M$ become spherical harmonics. For such a toy PTA where all pulsars have the same noise level, $w_i = w = \varepsilon/(1+\varepsilon)$ and
\begin{equation}
    \sum_i Y_{\alpha,i}^* Y_{\beta,i} \approx \frac{N_p}{4\pi}\delta_{\alpha\beta}.
\end{equation}
The Fisher matrices in Eqs.~\ref{eq:fisher_aa} and \ref{eq:fisher_bb} can then be rewritten as
\begin{align}
    F_{\alpha \beta} \approx&\ w\frac{N_p}{4\pi}\delta_{\alpha\beta},
    \label{eq:fisher_e} \\
    F^p_{\alpha \beta} \approx&\ w^2\frac{N_p}{4\pi}\delta_{\alpha\beta},
    \label{eq:fisher_p}
\end{align}
and the eigenvalues of $M$ are
\begin{equation}
    \lambda_{\ell} = \frac{2C_\ell}{N_\ell} = \frac{2C_\ell}{4\pi} w N_p,
    \label{eq:lambda_ell}
\end{equation}
each with multiplicity $2\ell + 1$. In other words, the spherical harmonics are the signal-to-noise eigenmodes and $\lambda_\ell$ are their eigenvalues. The total SNR$^2$ up to a maximum $\ell_{\rm max}$ therefore scales as $\rho^2_{e} (\ell_{\rm max}) \propto \sum_{\ell=2}^{\ell_{\rm max}} \tfrac{2\ell+1}{(\ell+2)(\ell+1)\ell(\ell-1)}$, suggesting that the map can be approximated by its first few multipoles with minimal loss of information. 

We can use the expression above to estimate which multipoles are measurable for a given toy-PTA. To reach an SNR$=1$ for $\ell=2,3,$ and $4$, we require $w N_p=4,20$, and $60$, respectively, noting that the maximum value of $w$ is 1 for an ideal noiseless pulsar. Hence, it is generally challenging to measure the GW signal beyond $\ell_{\rm max} \sim 2$. In a more realistic scenario, the sky locations and noise distribution of the pulsars lead to different uncertainties for each $\ell$ and $m$ with non-zero correlations between them, and the spherical harmonic basis no longer diagonalizes the SNR matrix. We discuss these points in more detail in Sec.~\ref{sec:sens} and \ref{sec:harm}.

We also highlight that the signal-to-noise ratio given by Eq.~\ref{eq:rho2_e} is the Earth term contribution, with the pulsar term included in the noise. While the pulsar term is often included and marginalized over in continuous wave searches, it is not typically included in the expression for the SNR. In the low SNR regime, we have that $\rho^2_{i,e} \approx |z_{i,e}|^2/\sigma^2_{i,n}$, while in the high SNR regime, $\rho^2_{i,e} \approx 2\left(1-2\sigma^2_{i,n}/|z_{i,e}|^2\right)$. Hence, the presence of the pulsar term modifies the dependence of $\rho^2_e$ as a function of the sky position. This is illustrated in Fig.~\ref{fig:SNR_EP}, where we evaluate Eq.~\ref{eq:rho2_e} assuming equal noise across pulsars/pixels. The left panel shows $\rho^2_{i,e}$ for $\varepsilon=0.01$ and the right panel for $\varepsilon=1$, showing the saturation of the SNR map due to the presence of the pulsar term. The pulsar term also contributes to the signal. We can define the SNR of the pulsar term map in the same way as Eq.~\ref{eq:rho2_e}, leading to
\begin{equation}
    \rho^2_p(f) = \sum_i \left(\frac{\sigma^2_{i,p}}{\sigma^2_i}\right)^2.
    \label{eq:rho2_p}
\end{equation}

\subsection{Fisher matrix for known templates}\label{sec:known_templates}
While in the previous section we considered the signal-to-noise of the Earth and pulsar term maps in general, one is typically interested in searching for signals with particular templates. The detection significance of a given signal depends on how it enters the likelihood. For a Gaussian likelihood with mean $\boldsymbol\mu(\theta)$ and covariance $C(\theta)$, the Fisher matrix is in general given by
\begin{equation}
F_{\theta\theta'} = \partial_\theta\boldsymbol\mu^\dagger\,C^{-1}\,\partial_{\theta'}\boldsymbol\mu
 + \Tr\!\left[C^{-1}\partial_\theta C\,C^{-1}\partial_{\theta'} C\right].
\label{eq:F_general_gauss}
\end{equation}
We will consider an amplitude parameter that scales a mean template (typically corresponding to a deterministic point source), $\boldsymbol\mu(A_s)=A_s \mathbf t$, and an amplitude that scales a covariance template (corresponding to a stochastic GW signal), $C(A)=A T + N$. Eq.~\ref{eq:F_general_gauss} then reduces to
\begin{align}
F_{A_s A_s} =& \mathbf t^{\,\dagger} C^{-1} \mathbf t,\qquad \label{eq:F_A_mean} \\
F_{AA} =& \Tr \left[C^{-1} T\,C^{-1} T\right]. \label{eq:F_A_cov}
\end{align}

\paragraph*{Stochastic backgrounds.} A stochastic signal is described by its covariance matrix $C$. Here we assume that it includes a HD-correlated component, a common uncorrelated red-noise (CURN) component, and a noise term. That is,
\begin{equation}
    C = A_{\rm H} \Gamma + A_{\rm C} I + N,
    \label{eq:C_k}
\end{equation}
where $A_{\rm H}$ is a normalized amplitude for the HD component, with a fiducial value of $A_{\rm H} = 1$, and $A_{\rm C}$ of a CURN term, with an assumed fiducial value of $A_{\rm C} = 0$. The Hellings and Downs covariance is given by $\Gamma_{ij} = \Gamma^{e}_{ij} + \delta_{ij}$, where the Earth term contribution $\Gamma^{e}_{ij}$ is related to $C_{\ell}$ via
\begin{equation}
    \Gamma^{e}_{ij} = \sum_{\ell=2} C_\ell \frac{2\ell+1}{4\pi} P_\ell(\hat n_i \cdot \hat n_j),
\end{equation}
and $P_\ell$ are the Legendre polynomials. The noise covariance matrix is $N_{ij} = \delta_{ij} \tfrac{\sigma^2_{n,i}}{\sigma_{h, {\rm fid}}^{2}} = \delta_{ij} \varepsilon_i^{-1}$. In the forecasts performed in this work, we assume that all pulsar-specific noise parameters and the spectral index of the GW signal are fixed, and we will estimate the sensitivity to an overall amplitude of the HD-correlated component. 

The Fisher matrices for $A_{\rm H}, A_{\rm C}$ are computed from Eq.~\ref{eq:F_A_cov}, with partial derivatives $T_{\rm H} = \partial_{A_{\rm H}} C = \Gamma$ and $T_{\rm C} = \partial_{A_{\rm C}} C = I$. Assuming that the timing residuals are measured at $N_f$ independent frequency bins labelled by $k$ and that the spectral index is fixed, the total Fisher matrix is the sum $F_{\theta \theta'} = \sum_{k=1}^{N_f} F_{\theta \theta'}(f_k)$. When including the CURN component, we compute $F_{A_{\rm H} A_{\rm H}}$, $F_{A_{\rm H} A_{\rm C}}$, and $F_{A_{\rm C} A_{\rm C}}$ from Eq.~\ref{eq:F_A_cov}, and the uncertainty on the HD amplitude marginalized over CURN is
\begin{equation}
    F'_{A_{\rm H} A_{\rm H}} = F_{A_{\rm H} A_{\rm H}} - \frac{F_{A_{\rm H} A_{\rm C}}^2}{F_{A_{\rm C} A_{\rm C}}}.
    \label{eq:marg_HH}
\end{equation}
Throughout this work, the CURN-marginalized Fisher matrix is denoted by a prime, $F'$.

\paragraph*{Deterministic sources.}
For a single source the same amplitude $A_s$ scales both the Earth-term mean, $\boldsymbol\mu=A_s \mathbf z_e$, and the pulsar-term variance, $\sigma^2_{i,p}\to A^2_s \sigma^2_{i,p}$. Both terms of Eq.~\ref{eq:F_general_gauss} therefore contribute. With $\partial_{A_s} \boldsymbol\mu=\mathbf z_e$ and $\partial_{A_s} C = 2A_s \mathrm{diag}(\sigma^2_{i,p})$, evaluating at the fiducial $A_{s}=1$ results in
\begin{equation}
F_{A_s A_s} = \sum_i \frac{|z_{e,i}|^2}{\sigma_i^2} + 4\sum_i\left(\frac{\sigma^2_{i,p}}{\sigma_i^2}\right)^{2}
 = \rho_e^2 + 4\rho_p^2,
\label{eq:F_AA_det}
\end{equation}
where the factor of 4 is due to the chain rule from $\sigma^2_p\propto A^2_s$. A CURN component adds a monopole to the pulsar-term variance with amplitude $A_{\rm C}$. Eq.~\ref{eq:marg_HH} for the deterministic source case gives
\begin{equation}
F'_{A_s A_s} = \rho_e^2 + 4\left[\rho_p^2
 - \frac{\big(\sum_i w_i^2\,\sigma^2_{i,p}/\sigma^2_h\big)^2}{\sum_i w_i^2}\right].
\label{eq:F_A_det_marg}
\end{equation}
Marginalizing over the CURN component leaves only the information in $\ell\geq 1$ multipoles of the pulsar-term map. For identical, uniformly distributed pulsars this leaves $F'_{A_s A_s} = \rho_e^2 + \tfrac{16}{9}\rho_p^2$, i.e. $4/9$ of the pulsar-term information survives.

\subsection{Realistic PTA}\label{sec:sens}

\begin{figure*}
    \centering
    \includegraphics[width=0.8\linewidth]{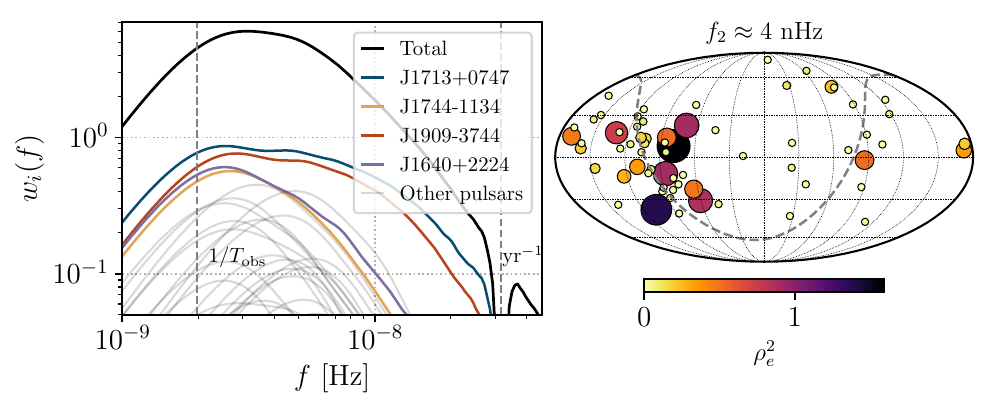}
    \caption{Sensitivity and sky locations for pulsars in the NANOGrav 15yr data set. The left panel shows the ratio between the GW power to the noise power ($w_i = \sigma^2_h/\sigma^2_{i, \mathcal{N}}$) as a function of frequency for each pulsar, with the first few most sensitive pulsars in color, and the sum over all pulsars in black. The right panel shows the location of the pulsars on the sky, with the size and colors indicating the SNR$^2$ of the Earth-term at the most sensitive frequency bin of $f_2 \equiv 2/T_{\rm obs} = 4\times 10^{-9}$ Hz.}
    \label{fig:SNR_NG15yr}
\end{figure*}

In order to ground the discussion in the current observational status of PTAs and provide realistic estimates for the map and HD sensitivity as a function of measured multipole and frequency, we consider the pulsars included in the NANOGrav 15yr analysis~\cite{NANOGrav_stoc} and will use their inferred noise properties to compute some of the sensitivity estimates discussed in Sec.~\ref{sec:FIM}. The noise estimates are computed using the \texttt{hasasia} package and the pulsar noise properties are obtained from the publicly available NANOGrav 15yr results~\cite{NANOGrav_15yr_dataset}. 

For a single pulsar $i$, subtracting the expected from the observed pulse arrival time and linearizing the timing model around its best-fit parameters, yields the timing residual
\begin{equation}
    \delta t_i(t) = M_i(t) \epsilon_i + n_i^{\rm WN}(t) + n_i^{\rm RN}(t) + s_i^{\rm GW}(t)
\end{equation}
where $M$ is the design matrix for the deterministic timing model, $\epsilon$ are the parameter offsets,  $n^{\rm WN}$ and $n^{\rm RN}$ are the white measurement noise and intrinsic red noise, respectively, and $s^{\rm GW}$ is the GW signal. The timing residual is related to the redshift by
\begin{equation}
    \delta t_i(t) = \int_0^t dt' z_i(t').
    \label{eq:delta_t_z}
\end{equation}
The timing model captures the dependence on the pulsar's spin, astrometry, binary motion, line-of-sight dispersion measure, among others~\cite{NANOGrav_noise}. Fitting the timing model is equivalent marginalizing the likelihood over $\epsilon$ assuming flat improper priors. This is typically computed by assuming a zero-mean Gaussian prior with variance $\phi_{\rm M}$ and taking $\phi_{\rm M} \to \infty$ \cite{NANOGrav_methods}.

For each unique receiver and backend system $\mu$, the white-noise variance is block-diagonal in arrival time ($k$, $l$) and diagonal in receiver and backend system \cite{NANOGrav_noise}
\begin{equation}
    \sum_{\mu} N^{\rm WN}_{\mu, kl} = \sum_{\mu} \left(\sigma^{\rm WN}_{\mu, k}\right)^2 \delta_{kl} + J_\mu^2 \mathcal{U}_{kl}
    \label{eq:white_noise_cov}
\end{equation}
where $(\sigma^{\rm WN}_{\mu, k})^2 \equiv G_\mu\,(\sigma_{\mathrm{TOA},k}^2 + Q_\mu^2)$, the white noise parameters $G_\mu$, $Q_\mu$, and $J_\mu$ correspond to EFAC, EQUAD, and ECORR, respectively, and $\mathcal{U}_{kl}$ is a block-diagonal matrix with ones for arrival times in the same epoch and zeros otherwise. In forecasts, the noise correlated across epochs is typically neglected and the uncorrelated component is idealized as a constant, leading to a white noise power spectral density of $P^{\rm WN}(f) = 2\sigma^{2} \Delta t$ for an assumed uniform cadence $\Delta t$. While these are reasonable approximations for forecasts, we include all contributions to the noise covariance since the noise properties are taken directly from the NANOGrav analysis.

The intrinsic red noise and GW signals are represented by a truncated Fourier series, such that 
\begin{equation}
    \mathbf{n}_i^{\rm RN} + \mathbf{s}_i^{\rm GW} = F_i \mathbf{c}_i ,
    \label{eq:fourier_series}
\end{equation}
where $F_i$ is a matrix of $e^{2 \pi i f_k t_l}$ for each discrete frequency $f_k$ and arrival time $t_l$ and $c_{i,k} = c^{\rm RN}_{i,k} + c^{\rm GW}_{i,k}$ is the $k$-th fourier coefficient for the $i$-th pulsar and includes both the red noise and GW signal. If both the intrinsic red noise and GWs are assumed to be stochastic, their variance is
\begin{equation}
    \avg{c^*_{i,k} c_{i,k'}} = \Phi_k\delta_{kk'} = \left[ \varphi_{i,k} + \phi_k \Gamma_{ii} \right] \delta_{kk'},
\end{equation}
where $\varphi_{i,k} = P^{\rm RN}_{i}(f_k) \Delta f$ is the intrinsic red noise power spectral density and $\phi_k = P^{\rm GW}(f_k) \Delta f$ is the GW power, and $\Gamma$ is the overlap reduction function (ORF). Both red noise components are modelled as power-law spectra, with
\begin{equation}
    P(f) = \frac{A^2}{12\pi^2} \left(\frac{f}{f_{\mathrm{ref}}}\right)^{-\gamma}.
\end{equation}
From Eq.~\ref{eq:delta_t_z}, we can identify the fourier coefficient of the timing residual $c^{\rm GW}_{i,k}$ with the redshifts of Eq.~\ref{eq:d_i}: $z_{i,e}(f_k) + z_{i,p}(f_k) = 2\pi i f_k c^{\rm GW}_{i,k}$.

Marginalizing over the timing-model parameters $\epsilon_i$, white noise, intrinsic red noise, and the pulsar term GW leads to the time-domain likelihood
\begin{equation}
\ln p(\delta \mathbf{t}_i | \mathbf{s}^{\rm GW}_{i,e}) \propto -\frac{1}{2} (\delta \mathbf{t}_i - \mathbf{s}^{\rm GW}_{i,e})^T C_i^{-1} (\delta \mathbf{t}_i - \mathbf{s}^{\rm GW}_{i,e}),
\label{eq:lnlike_time_domain}
\end{equation}
where $C_i = N^{\rm WN}_i + M_i \phi_{\rm M} M_i^T + F_i \Phi F_i^T$ is the noise covariance matrix. Note that, at this stage, we may choose to marginalize only over the pulsar term as shown in the expression above, which results in $\Phi_k = \varphi_{i,k} + \phi_k/2$, or take the standard approach in stochastic GWB searches, which marginalizes over both the Earth and pulsar terms, resulting a zero-mean Gaussian and $\Phi_k = \varphi_{i,k} + \phi_k$. The flat-prior limit $\phi_{\rm M}\to\infty$ is well-defined for $C_i^{-1}$ via the Woodbury identity~\cite{NANOGrav_methods}. Changing to the Fourier basis $F_i$ yields a Gaussian likelihood for $\widetilde{\delta t}_i(f_k)$ 
\begin{equation}
\ln p(\widetilde{\delta t}_i | \tilde{s}^{\rm GW}_{i,e}) \propto -\frac{1}{2} (\widetilde{\delta t}_i - \tilde s^{\rm GW}_{i,e})^T \mathcal N_i^{-1} (\widetilde{\delta t}_i - \tilde s^{\rm GW}_{i,e}),
\label{eq:lnlike_freq_domain}
\end{equation}
where the mean is $\tilde s^{\rm GW}_{i,e}(f_k) = z_{i,e}(f_k)/(2\pi i f_k)$ (from Eq.~\ref{eq:delta_t_z}) and covariance
\begin{equation}
\mathcal N_i^{-1} \equiv \frac{1}{2T} F_i^T C_i^{-1} F_i,
\label{eq:N_freq}
\end{equation}
which in this case only includes the discrete frequencies $f_k$ of $F_i$, but can be considered in the continuous limit as in Ref.~\cite{Hazboun:2019vhv}. 

We can use these noise estimates to extend the results of Sec.~\ref{sec:FIM} to account for more realistic sources of noise and the marginalization over the timing model. Since the inverse covariance matrix $\mathcal N_i^{-1}$ is well-approximated as being diagonally dominated, the variance of the pulsar redshift in a finite frequency bin given in Eq.~\ref{eq:p_gauss} is related to $\mathcal{N}_i^{-1}$ via
\begin{equation}
    \frac{1}{\sigma^2_{i,\mathcal{N}}} = \frac{2}{(2\pi f)^2 \Delta f}\,\mathcal{N}^{-1}_i(f).
    \label{eq:N_i}
\end{equation}
The estimates in Sec.~\ref{sec:FIM} can be extended by taking $\sigma^2_{i} \rightarrow \sigma^2_{i,\mathcal{N}}$. In the regime where $\sigma^2_{i,\mathcal{N}} \approx (\sigma^2_{i,n} + \sigma^2_{i,p})/\mathcal{T}_i$, the pulsar's contribution to the SNR$^2$ is $\mathcal{T}_i/(1+\varepsilon_i^{-1})$, and we can therefore see that the timing model lowers the effective number of pulsars, such that the resulting SNR contribution of a pulsar can be $<1$ even in the absence of noise. 

The same substitution extends the HD and CURN Fisher matrix calculation in Eq.~\ref{eq:F_general_gauss} to the more realistic case. Starting again from Eq.~\ref{eq:lnlike_freq_domain} and marginalizing over the Earth term recovers the usual PTA likelihood. The derivative of the covariance with respect to $A_{\rm H}$ and $A_{\rm C}$ is unchanged, i.e. $\partial_{A_{\rm H}} C = \Gamma$ and $\partial_{A_{\rm C}} C = I$, while the covariance in the fiducial model and assuming the diagonal-in-frequency approximation is
\begin{equation}
\left[ C_k^{\rm fid} \right]_{ij} = \delta_{ij}\frac{\sigma^2_{i,\mathcal N}(f_k)}{\sigma^{{\rm fid},2}_h(f_k)} + \Gamma^{e}_{i\neq j},
\label{eq:C_k_realistic}
\end{equation}
where $\sigma^2_{i,\mathcal N}$ includes both the Earth and pulsar term auto-correlations. 

To obtain a realistic prediction of the effective weights $w_i(f)$, we construct the time-domain noise covariance for each pulsar using the best-fit values from NANOGrav 15yr analysis~\cite{NANOGrav_15yr_dataset}. This includes the timing model marginalization, white noise, and both intrinsic and GW-induced red noise, with the fiducial GWB spectrum given by $A_h = 6.4 \times 10^{-15}$ and $\gamma_h = 3.2$. Fig.~\ref{fig:SNR_NG15yr} shows the weights $w_i$ as a function of frequency for each pulsar, with the first few most sensitive pulsars in color, and the sum over all pulsars in black. The left panel shows that, for a single frequency bin, the maximum effective number of pulsars achieved peaks around $N^{\rm eff}_p \sim 6$ at $f_2 \equiv 2/T_{\rm obs} = 4\times 10^{-9}$ Hz. 

The panel on the right shows the location of the pulsars on the sky, with the size and colors indicating the map SNR$^2$ of the Earth-term $\rho^2_e$ at the most sensitive frequency bin $f_2$, highlighting the non-uniformity of the weight distribution on the sky. While the left panel makes it clear that the relative importance of each pulsar can vary significantly with frequency, it is generally true that the weights are dominated by a few pulsars across all frequencies. At $f=4$nHz, half of the map SNR$^2$ is contributed by the first 6 pulsars, while at and above $f=10$nHz, the 2 most sensitive pulsars dominate.

\begin{figure}[t]
    \centering
    \includegraphics[width=0.95\linewidth]{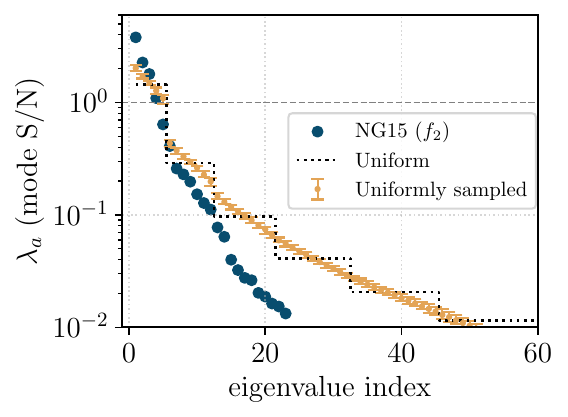}
    \caption{Eigenvalue spectrum of the whitened Earth-term signal covariance matrix $M=W^{1/2} S W^{1/2}$ (Eq.~\ref{eq:snr_matrix}), evaluated at the most sensitive NG15 frequency bin $f_2\approx4$~nHz. Each eigenvalue $\lambda_a$ is the SNR$^2$ of an independently measurable mode of the Earth-term map, and their sum is the total $\rho^2_e=\Tr M$ (Eq.~\ref{eq:rho2_e}). Blue filled circles show the realistic NG15 array (per-pulsar weights $w_i$ and true sky locations). The yellow error bars show the median and 16th--84th percentile over 100 random draws of $N_p=67$ uniformly distributed, equal-weight pulsars. The dotted line is the analytic uniform-array prediction given by Eq.~\ref{eq:lambda_ell}. The dashed line marks $\lambda=1$, above which a mode is signal-dominated.}
    \label{fig:eigs}
\end{figure}

Fig.~\ref{fig:eigs} shows the signal-to-noise eigenvalues discussed in Sec.~\ref{sec:FIM} of the matrix $M$ \ref{eq:snr_matrix} in the ideal diagonal Fisher matrix approximation from Eq.~\ref{eq:lambda_ell}, a case where all pulsars have identical noise properties and are uniformly distributed on the sky, and the realistic noise estimates for the NG15 with the weights $w_i$ evaluated at the most sensitive frequency bin $f_2$. In all three cases, the total SNR (given by the trace of $M$) is the same, and the comparison between the blue dots and the dotted line shows the non-uniform noise properties of a realistic PTA results in a smaller number of signal-dominated modes.

\section{How to make PTA maps without losing information}\label{sec:harm}

The maps of the Earth term and pulsar term variance contain all of the information about GW sources available to PTAs. Since one or a superposition of any number of sources produces, on average, an Earth term map with the same Hellings and Downs power spectrum (Eq.~\ref{eq:cl_ze}) which is a steeply decaying function of $\ell$, it can be accurately described by a small set of spherical harmonic coefficients. We can therefore truncate the sum over modes in Eq.~\ref{eq:loglike} at a small $\ell_{\max}$ while retaining most of the information. For uniform/equal-noise pulsars each mode is measured independently with the SNR scaling of Sec.~\ref{sec:FIM}, but the realistic noise and sky distribution of Sec.~\ref{sec:sens} couples the modes and gives them unequal errors. 

In this section we quantify how much information about stochastic or deterministic point sources is lost by compressing to $\ell\le \ell_{\max}$. We consider a realistic setting, in which the pulsar locations and noise properties are obtained from the NG15 dataset~\cite{NANOGrav_15yr_dataset}. In particular, we will consider the mean or covariance templates introduced in Sec.~\ref{sec:FIM} and parametrize their amplitudes $A$, as in Eq.~\ref{eq:F_A_mean} and \ref{eq:F_A_cov}. How much information the compression retains about the relevant signal is quantified by the ratio $\mathcal{R}$ between the Fisher matrix with respect to $A$ for the compressed statistic ($\tilde{F}$) and the full observable ($F$). That is, we evaluate the ratio
\begin{equation}
    \mathcal{R}^2 \equiv \frac{\tilde{F}_{AA}}{F_{AA}},
    \label{eq:R_general}
\end{equation}
or $\mathcal{R}^{' 2} = \frac{\tilde{F}'_{AA}}{F'_{AA}}$ for the CURN-marginalized case. Below we derive the mean and covariance of the observed spherical harmonic coefficients required to compute $\tilde{F}_{AA}$, before moving on to the Fisher matrix ratios for stochastic and deterministic point-source signals.

\subsection{Compressed statistic and its moments}
Starting from Eq.~\ref{eq:loglike}, the maximum-likelihood estimator for $\mathbf{a}$ is derived by taking $\partial\ln p/\partial a^*_\alpha=0$ at fixed pulsar term $b$, which yields the usual solution
\begin{align}
\hat{\mathbf{a}} = F^{-1} \tilde{\mathbf{a}}, \quad \text{where} \quad \tilde a_\alpha = \sum_i w_i Y^*_{\alpha,i} \frac{d_i}{\sigma_h},
\label{eq:alm_MLE}
\end{align}
and $Y$ is the $N_p\times M$ design matrix for $(\ell_* +1)^2$ spherical harmonics and $\ell_*$ is a large $\ell$ cutoff. As in Sec.~\ref{sec:FIM}, the matrix $W$ contains the inverse-variance weights, and $F = Y^{\dagger} W Y$ is the Fisher matrix, which is deconvolved in the estimator $\hat{\mathbf{a}}$.

The observed (``dirty'') and deconvolved (``clean'') coefficients $\tilde a$ and $\hat a$ are related by a linear transformation, and therefore have the same Fisher information. When $F$ is ill-conditioned, it must be regularized and the clean estimator is obtained by computing the pseudo-inverse. For this reason, it is typically more convenient to work directly in terms of the observed map $\tilde a$. Compressing to $\ell \le \ell_{\max}$ means keeping only the low-$\ell$ ($L$) block $\tilde{\mathbf a}_L = Y_L^\dagger W\mathbf d/\sigma_h$, with the remaining high-$\ell$ ($H$) modes discarded. All subsequent calculations depend on the data only through the first two moments of the retained observables, which we now derive. We consider two cases: moments averaged over an isotropic stochastic signal, used in the stochastic subsection, and moments conditioned on a realization of the Earth-term map $a_\alpha$, which will be used for the deterministic point-source signals.

Starting with the Earth term estimator, we have that the first two moments of the observed harmonic coefficients, under the data model in Eq.~\ref{eq:d_i} and conditioned on a realization of the signal, are
\begin{equation}
\avg{\tilde a_\alpha} = \sum_\beta F_{\alpha\beta}\,a_\beta,\qquad \avg{\tilde a^*_{\alpha'}\tilde a_\alpha}-\avg{\tilde a^*_{\alpha'}}\avg{\tilde a_\alpha} = 2 F_{\alpha\alpha'}.
\label{eq:alm_moments}
\end{equation}
The moments of the deconvolved estimator are $\avg{\hat a_\alpha}=a_\alpha$ and $\avg{\hat a^*_{\alpha'}\hat a_\alpha}-\avg{\hat a^*_{\alpha'}}\avg{\hat a_\alpha}=2\,[F^{-1}]_{\alpha\alpha'}$. Restricting Eq.~\ref{eq:alm_moments} to the retained block gives the compressed moments. Note that the sum over $\beta$ in the mean includes all $\ell$'s, which gives the compressed mean an aliasing contribution: $\avg{\tilde{\mathbf{a}}_L} = F_L \mathbf{a}_L + F_{LH}\mathbf{a}_H$, where $F_L=Y_L^{\dagger}WY_L$ is the low-$\ell$ block of the coupling matrix and $F_{LH}=Y_L^{\dagger}WY_H$ couples the low- and high-$\ell$ blocks. This leakage vanishes only in the uniform-pulsar limit, $F_{LH}\to0$.

For a stochastic signal, we must also consider the signal covariance. Including the Earth and pulsar terms and noise contributions to $d_i$ in \ref{eq:alm_MLE}, we find that the covariance per real/imaginary component of the observed Earth term map is
\begin{equation}
    \tilde{C}_{\alpha\alpha'} = \sum_\gamma C_{\ell_\gamma} F_{\alpha\gamma} F_{\gamma\alpha'} + F_{\alpha\alpha'},
    \label{eq:tildeC_alpha}
\end{equation}
where $C_\ell$ is given by Eq.~\ref{eq:cl_ze}. 

The pulsar-term variance map is estimated analogously by the quadratic estimator
\begin{equation}
    \hat{\mathbf{b}} = \left(F^{p}\right)^{-1} \tilde{\mathbf{b}}, \quad \text{where} \quad \tilde b_\alpha = \sum_i w_i^2\, Y^*_{\alpha,i} \frac{u_i}{\sigma_h^2},
    \label{eq:blm_MLE}
\end{equation}
where $u_i$ is built from the residual $r_i$ between the data and the earth term map, and is defined as
\begin{equation}
    u_i = \tfrac{1}{2} \left| r_i \right|^2 - \sigma_{n,i}^2.
    \label{eq:u_i_def}
\end{equation}
We make slightly different assumptions in the definition of $r_i$ in the deterministic and stochastic cases. In the deterministic case, we assume that the residual is constructed at the fiducial template, hence $r_i = d_i - \sigma_h\sum_{\alpha} a_\alpha Y_{\alpha,i}$. The moments of $\tilde{\mathbf{b}}$ then follow the same pattern as Eq.~\ref{eq:alm_moments}, with
\begin{equation}
    \langle\tilde b_\alpha\rangle = \sum_\beta F^p_{\alpha\beta}\,b_\beta,\qquad \langle\tilde b^*_{\alpha'}\tilde b_\alpha\rangle - \langle\tilde b^*_{\alpha'}\rangle \langle\tilde b_\alpha\rangle = F^p_{\alpha\alpha'}.
    \label{eq:blm_moments_det}
\end{equation}

In the stochastic case, we assume that the residual is evaluated at the fitted template, such that $r_i = d_i - \sigma_h\sum_{\alpha \le \ell_{\max}}\hat a_\alpha Y_{\alpha,i}$, where the fit runs over the same retained block as the compression. Substituting $\hat{\mathbf a}_L = F_L^{-1}Y_L^\dagger W \mathbf d/\sigma_h$ shows that the residual is a fixed linear filter of the data, $\mathbf{r} = P \mathbf{d}$, with the projector
\begin{equation}
    P = I - Y_L F_L^{-1} Y_L^{\dagger} W,
    \label{eq:P_def}
\end{equation}
which annihilates what the fitted modes can represent and leaves a residual orthogonal to the fit in the inverse-variance inner product. For an isotropic stochastic signal the moments of the compressed pulsar-term map are
\begin{equation}
\begin{split}
    \langle \tilde b_\alpha \rangle =&\ \sum_i w_i^2\, Y^*_{\alpha,i}\left[(PCP^\dagger)_{ii} - \frac{\sigma^2_{n,i}}{\sigma_h^2}\right], \\
    \langle \tilde b^*_{\alpha'} \tilde b_\alpha \rangle - \langle \tilde b^*_{\alpha'} \rangle \langle \tilde b_\alpha \rangle =&\ \sum_{ij} w_i^2 w_j^2\, Y^*_{\alpha,i} Y_{\alpha',j}\, \big|(PCP^\dagger)_{ij}\big|^2,
\end{split}
\label{eq:blm_moments_stoc}
\end{equation}
where $C$ is the full covariance of Eq.~\ref{eq:C_k}. The response and variance of the pulsar-term monopole used in Sec.~\ref{sec:stoc} are the $\alpha=(00)$ components of Eq.~\ref{eq:blm_moments_stoc} and of its derivative with respect to the amplitudes.

\subsection{Stochastic signals}\label{sec:stoc}
For a stochastic signal, the parameters of interest are the amplitudes $A_{\rm H}$ and $A_{\rm C}$ of the HD and CURN covariance templates of Eq.~\ref{eq:C_k}. From Eq.~\ref{eq:tildeC_alpha}, the derivatives of the compressed Earth-map covariance are
\begin{equation}
    \partial_{A_{\rm H}} \tilde C_{\alpha\alpha'} = \sum_\gamma C_{\ell_\gamma}\, F_{\alpha\gamma}F_{\gamma\alpha'} + F^p_{\alpha\alpha'}, \qquad \partial_{A_{\rm C}} \tilde C_{\alpha\alpha'} = F^p_{\alpha\alpha'},
\end{equation}
with the coefficients restricted to $\ell_\alpha, \ell_{\alpha'} \le \ell_{\max}$ for the compressed statistic. The term $F^p$ in the first derivative arises because the same amplitude parametrizes both the Earth- and pulsar-term covariances, while the CURN component contributes uniformly to the variance of each pulsar and therefore contributes through the same matrix as the pulsar term. The Fisher matrices $\tilde F_{AA}$ and $F_{AA}$ follow from the covariance case of Eq.~\ref{eq:F_A_cov}, summed over frequency bins, and the CURN-marginalized version $F'$ follows from Eq.~\ref{eq:marg_HH}.

Along with the Earth-term map, one can measure the monopole of the pulsar-term variance, $\tilde b_{00}$, which carries the auto-correlation information. The response and the variance of $\tilde b_{00}$ are the $\alpha = (00)$ components of Eq.~\ref{eq:blm_moments_stoc},
\begin{equation}
\begin{split}
\partial_\theta \langle \tilde b_{00}\rangle =& \frac{1}{\sqrt{4\pi}}\sum_i w_i^2\, \big(P\, \partial_\theta C\, P^\dagger\big)_{ii}, \\ \sigma^2(\tilde b_{00}) =& \frac{1}{4\pi}\sum_{ij} w_i^2 w_j^2\, \big|(PCP^\dagger)_{ij}\big|^2,
\end{split}
\end{equation}
where the covariance $C$ and its derivatives $\partial_\theta C$ are the same as in Sec.~\ref{sec:known_templates}. Since $\tilde{\mathbf a}_L$ and $\tilde b_{00}$ are statistically independent, their Fisher information adds, $F_{\theta\theta'} = F^a_{\theta\theta'} + F^b_{\theta\theta'}$, with $F^b_{\theta\theta'} = \partial_\theta\langle\tilde b_{00}\rangle\, \partial_{\theta'}\langle\tilde b_{00}\rangle / \sigma^2(\tilde b_{00})$.

\begin{figure}[t]
    \centering
    \includegraphics[width=0.45\textwidth]{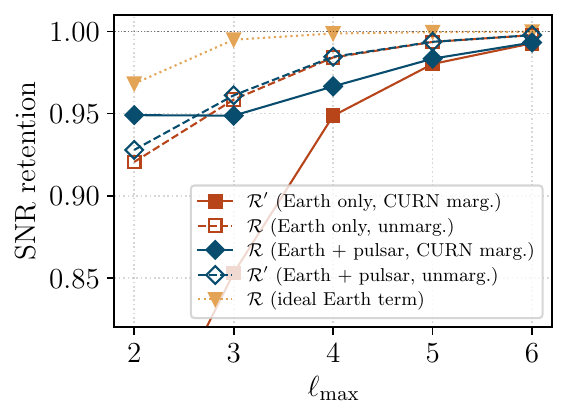}
    \caption{Fraction of the SNR of a stochastic GWB amplitude that is retained when the Earth term map is truncated to the first $\ell_{\rm max}$ multipoles. Unfilled and filled symbols ($\mathcal{R}/\mathcal{R}'$) correspond to the cases unmarginalized and marginalized over the CURN amplitude, respectively. The yellow triangles are the ideal Earth-term only result in the weak signal approximation, identical to the result in Ref.~\cite{Roebber:2016jzl}, while the other symbols use the realistic noise values for the NANOGrav 15yr (NG15) dataset. Red squares show the results including only the Earth term information, while blue diamonds include the pulsar term information as well.}
    \label{fig:HD_fisher_RL_compression}
\end{figure}

Fig.~\ref{fig:HD_fisher_RL_compression} shows the ratio $\mathcal{R}$ between the compressed and uncompressed Fisher matrices for the HD component of a stochastic GWB, summed over frequencies in the NG15 dataset. The yellow triangles and filled and open squares only include the Earth term information $\tilde{a}$, while the blue diamonds include the pulsar term information as well. The filled red squares and blue diamonds show results marginalized over the CURN amplitude. The gap between the red squares (unmarginalized and marginalized over the CURN) is due to the degeneracy between the auto- and cross-correlation patterns that is produced by the $\ell_{\rm max}$ truncation. If only a small number of harmonics are kept, an $\ell=2$ angular correlation pattern or a pure auto-correlation (flat in $\ell$) will appear similar, and including more modes helps to distinguish between the two. Once the pulsar term is included, this constrains the auto-correlation and the gap closes. The blue curve shows that a measurement of $\ell=2$ multipoles of the Earth term map and the monopole of the pulsar term variance is approximately equivalent to a measurement of the full PTA data set.

Note that $\mathcal{R}$ quantifies how much information the compression loses about a fixed signal template (an overall amplitude), not how well each mode is independently measured. A less uniform PTA generally has a higher $\mathcal{R}$, since the red $C_\ell$ spectrum concentrates the dominant SNR eigenmodes at low $\ell$ and the first few spherical harmonics span essentially the same space as the well-measured eigenmodes, even though for a non-uniform array the harmonics are not themselves the eigenmodes. The high-$\ell$ signal in a non-uniform array is also degenerate with the retained low-$\ell$ modes and aliases into them through $F_{LH}$, so it is recaptured rather than lost.

\subsection{Deterministic point-source signals}
In the case of a deterministic signal, the derivatives of the signal template with respect to a normalized source amplitude $A_s$ are
\begin{equation}
    \partial_{A_s}\avg{\tilde{\mathbf a}} =\ F\mathbf a, \quad \text{and} \quad \partial_{A_s}\avg{\tilde{\mathbf a}_L} =\ F_L \mathbf{a}_L + F_{LH}\mathbf{a}_H.
\end{equation}
From Eq.~\ref{eq:alm_moments}, the covariance matrices are $F$ and $F_L$, respectively. The matched filter SNR$^2$ in the compressed basis is therefore
\begin{equation}
    \begin{split}
    \tilde F_{A_s A_s} \equiv \tilde{\rho}^2_e =& \ \partial_{A_s}\langle \tilde{\mathbf a}_L \rangle^{\dagger}\, F_L^{-1}\, \partial_{A_s}\langle \tilde{\mathbf a}_L \rangle, \\
    =& \sum_{\alpha,\alpha'\le \ell_{\rm max}}\sum_{\beta,\beta'} [F_L^{-1}]_{\alpha'\alpha}\, F_{\alpha\beta}\, F^*_{\alpha'\beta'}\, a^*_{\beta'} a_\beta
\label{eq:tilderho_e}
\end{split}
\end{equation}
where $\beta,\beta'$ run over all harmonics, and in the second line we simply write out the indices explicitly. Similarly for the pulsar-term variance map, we have that 
\begin{equation}
    F^{b}_{A_s A_s} = 4 (F^p \mathbf{b})_L^{\dagger} (F^p_L)^{-1} (F^p \mathbf{b})_L,
\end{equation}
where, as in Eq.~\ref{eq:F_AA_det}, the factor of 4 arises from the fact that $b_{\ell m} \propto A^2_s$. The subscript $L$ in this case denotes the low-$\ell$ block for the pulsar term, which is generally defined by a different $\ell_{\rm max}$ choice.

We also consider the marginalization over the CURN amplitude $A_{\rm C}$. From Eq.~\ref{eq:blm_moments_det}, the derivative of the pulsar-term variance with respect to $A_{\rm C}$ is $\partial_{A_{\rm C}}\langle\tilde b_\alpha\rangle = \sqrt{4\pi} F^p_{\alpha,00}$, the monopole column of $F^p$ (using $Y_{00}=1/\sqrt{4\pi}$). Similarly to Eq.~\ref{eq:F_AA_det}, marginalizing via Eq.~\ref{eq:marg_HH} removes only the monopole projection of the pulsar-term,
\begin{equation}
F'_{AA} = \tilde\rho_e^2 + 4\left[(F^p\mathbf b)_L^\dagger (F^p_L)^{-1}(F^p\mathbf b)_L
 - \frac{\big|(F^p\mathbf b)_{00}\big|^2}{F^p_{00,00}}\right],
\label{eq:F_AA_det_marg_compressed}
\end{equation}
which is the analogue of Eq.~\ref{eq:F_A_det_marg} in the compressed basis. As in Sec.~\ref{sec:stoc}, we add the Fisher matrices for the Earth and pulsar term maps, and compute the retention ratios of Eq.~\ref{eq:R_general} for both marginalized over CURN and unmarginalized cases.

We consider a polarization and inclination averaged template for simplicity. The full single-source template for a binary in the $\hat z$ direction is given in Eq.~\ref{eq:alm_1source}, and produces a contribution $a^*_{\ell m} a_{\ell'm'} = (z_\ell z_{\ell'}/4\sigma_h^2) h^*_{m} h_{m'}$ for $m,m'=\pm 2$, and is zero otherwise, to the Fisher matrix in Eq.~\ref{eq:tilderho_e}. Averaging over the source phase $\Phi_0$, polarization angle $\psi$, and inclination $\cos\iota$ at fixed intrinsic amplitude diagonalizes across $m,m'$ and sets $\langle|h_2|^2\rangle = \langle|h_{-2}|^2\rangle$. The averaged template reduces to
\begin{equation}
    \langle a^{*}_{\ell m} a_{\ell'm'}\rangle_{\Phi_0, \psi, \cos\iota} = 3 z_\ell z_{\ell'}\,\delta_{mm'}\,(\delta_{m,2}+\delta_{m,-2}),
\end{equation}
noting that they do not diagonalize across $\ell$ as it would in the isotropic stochastic case. The harmonic coefficients for a source at a general direction $\hat\Omega_s$ is obtained by rotating the sky coordinates from $\hat{z}$, that is, 
\begin{equation}
    a_{\ell m}(\hat\Omega_s) = \sum_{m'} D^{\ell}_{m m'} (\phi_s,\theta_s, 0) a_{\ell m}(\hat{z}), 
\end{equation}
where $D^{\ell}_{m m'}$ are the Wigner D-matrix elements. Note that, when averaged over polarizations, the pulsar term template becomes non-zero only for $\ell \leq 2$. Given the hierarchy in coefficients derived in Eqs.~\ref{eq:bLM}, we only include $\ell = 0, 1$ in the forecast.

\begin{figure*}[t]
    \centering
    \includegraphics[width=0.8\textwidth]{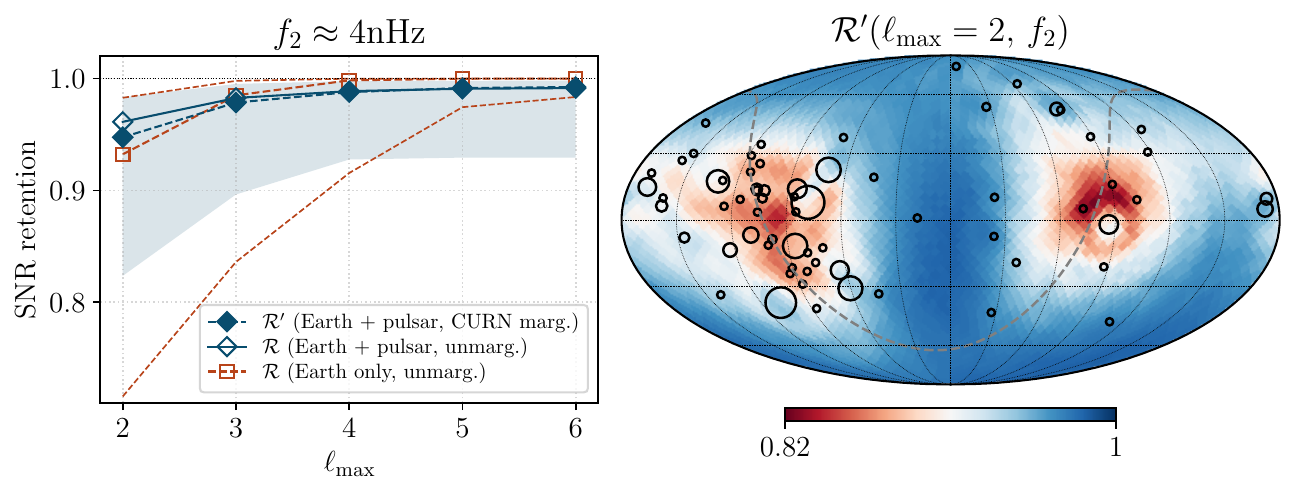}
    \caption{Fraction of the SNR that is retained when the Earth term map is only measured up to a maximum multipole $\ell_{\rm max}$, assuming a NG15-like PTA at the most sensitive frequency ($f_2 \approx 4$~nHz). The left panel shows the median, and minimum and maximum band across the sky of the SNR retention $\mathcal{R}$~\eqref{eq:R_general} as a function of $\ell_{\rm max}$. The red squares corresponds to the fraction of the SNR of a single source amplitude retained considering only the Earth term information, while the unfilled and filled blue diamonds show the fraction including both the Earth and pulsar term, unmarginalized and marginalized over a CURN amplitude ($\mathcal{R}$/$\mathcal{R}'$), respectively. The panel on the right shows a sky map of $\mathcal{R}'$ for $\ell_{\rm max} = 2$, and the unfilled black circles correspond to the pulsar sky locations.}
    \label{fig:MF_src_R_L}
\end{figure*}

Fig.~\ref{fig:MF_src_R_L} shows $\mathcal{R}$ for a deterministic point source, computed for a source located at every pixel on the sky. The panel on the left shows the median and range of $\mathcal{R}$ values as the source location varies across the sky, as a function of the $\ell_{\rm max}$ truncation of the Earth term map. The plateau in the minimum of $\mathcal{R}'$ (blue band) is due to the fact that we only include $\ell = 0, 1$ in the pulsar term. We can see that, while the median SNR retention is large ($> 0.9$) for $\ell_{\rm max} = 2$, it can vary significantly across the sky. In particular, the sky locations most affected by the truncation are near the sensitive pulsars. This is due to the fact that the timing residuals produced by a single source have a large amplitude near the source location (in a cross pattern), which is produced by higher $\ell$ modes. While this reduces the SNR retention once we truncate in $\ell$, it remains true that the contribution falls with $\ell$ as described in Sec.~\ref{sec:EP_SH}. Hence, we still retain a large fraction of the information about the source even for $\ell_{\rm max} = 2$ at the worst sky location. An $\ell_{\rm max} \sim 3$ and the pulsar term up to the dipole are sufficient to achieve an SNR retention of $\geq 0.9$ across the entire sky, and is $\geq 0.8$ even for $\ell_{\rm max} = 2$.

\section{Conclusion}
Following the detection of a GW signal with PTAs, the driving question shifts to the characterization of the signal and its sources. Distinguishing an isotropic stochastic background from an anisotropic or point-source-dominated sky, and ultimately resolving individual supermassive black hole binaries is of particular interest for gravitational-wave astronomy. 

Any type of the aforementioned GW signals produces, on average, the same Hellings and Downs angular correlation. The fact that the Earth-term power spectrum falls steeply with $\ell$, as $C_\ell \propto 1/[(\ell+2)(\ell+1)\ell(\ell-1)]$ (Eq.~\ref{eq:cl_ze}), regardless of the nature of the source, suggests that the GW sky as seen by a PTA is inherently low-dimensional and can therefore be summarized in a small number of parameters. The main result of this work is to show that most of the information about GW signals is encoded in a small number of spherical harmonic coefficients of the Earth term map and the pulsar term variance map, which form a nearly lossless summary statistic for the GW content of PTA data. While the optimal compression is achieved by directly computing the eigenmodes of the whitened covariance matrix $M$~\eqref{eq:snr_matrix}, we focus on the spherical harmonic basis since it already provides a substantial compression of the data set, while remaining interpretable and independent of assumptions about the signal~\cite{Cornish:2014rva, Ali-Haimoud:2020iyz}.

We quantified this in a realistic forecast using the noise properties and sky locations of the NANOGrav 15yr pulsars. Retaining only the $\ell=2$ modes of the Earth-term map and the monopole of the pulsar-term variance preserves $\sim 95\%$ of the information about an isotropic background. Depending on the sky location, point sources may demand a few more modes due to the uneven sky distribution and sensitivity of pulsars. The dipole of the pulsar term variance contributes significant information about an individual source, and we show that including it along with the Earth term up to $\ell_{\rm max}\sim 3$ contains almost all of the information for sources at any sky location. Additionally, we note that since the need for an increase in the number of modes for the individual source is driven primarily by the non-uniformity of the pulsar sky distribution and noise, we may require fewer modes to capture most of the information as the number of pulsars increases and PTAs sample the sky in a more uniform manner.

We also highlight that each of the aforementioned analyses (isotropic, anisotropic, or continuous-wave searches) are a downstream operation on the proposed signal-agnostic summary. A practical upshot is that a single analysis over the full timing-residual dataset suffices to extract on the order $\sim$ten coefficients per frequency bin for the first lowest frequency bins that carry all of the signal. After this, searches for any signal or hypothesis testing can be carried out cheaply and consistently on the compressed summary.

\acknowledgments
We would like to thank Nick Kokron and Gil Holder for helpful discussions. GSP acknowledges support from the Friends of the Institute for Advanced Study Fund. MZ acknowledges support from the National Science Foundation NSF-BSF 2207583 and NSF 2209991, the Nelson Center for Collaborative Research and the Simons Foundation through the Black Holes and Strong Gravity program through Award No. SFI-MPS-BH-00012593-10. This work used large language models (LLMs) including Claude Fable 5 and Opus 4.8 in the later stages of the project with supervision by the authors.

\section*{Data Availability}
The code and pre-computed data products that reproduce all figures in this paper are publicly available~\cite{suff_stats_code}.

\bibliography{ref.bib}
\bibliographystyle{utcaps}

\appendix

\section{Gaussian approximation for the pulsar term}\label{app:gauss_approx}

\begin{figure*}[t]
    \centering
    \includegraphics[width=0.65\linewidth]{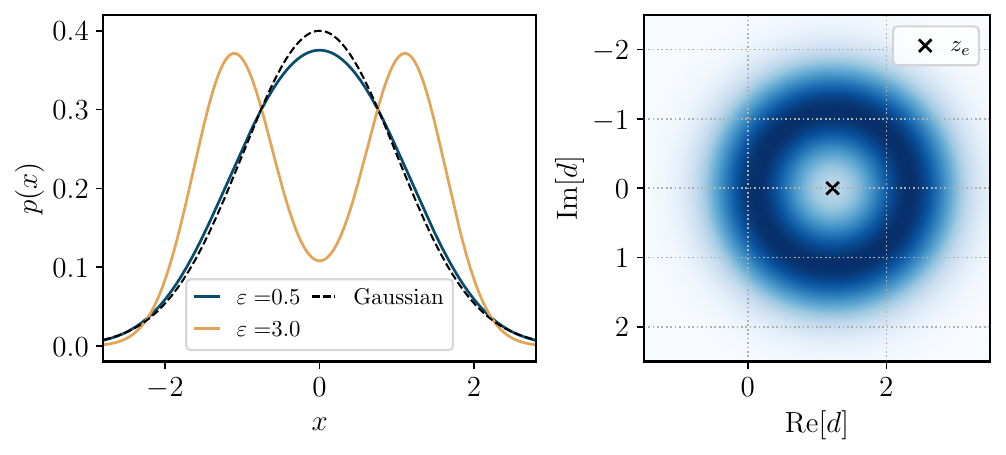}
    \caption{Distributions $p$ and $q$ for different values of $\varepsilon = |z_{e,1}|^2/(2\sigma_N^2)$, with $\sigma^2_N(1+\varepsilon) = 1$ held fixed. Left panel: cross-sections of $p$ and $q$ along the real axis, where $x=d-z_e$. Right panel: the distribution $q$ in the complex plane for $\varepsilon=5$, centered on $z_e \approx 1$.}
    \label{fig:pdfs}
\end{figure*}

In this appendix, we quantify the error introduced by adopting the Gaussian approximation to the likelihood given in Eq.~\ref{eq:p_gauss}. The derivation of Eq.~\ref{eq:p_gauss} assumed that $z_{i,p}$ follows a Gaussian distribution, which emerges in the limit of many sources. Here, we consider the opposite limit, in which the signal is dominated by a single GW source (i.e., $z_{i,p} = z_{i,e} e^{i\psi_{i}}$). In this case, the correct likelihood is obtained by marginalizing over the phase $\psi_{i}$, which we assume to be uniformly distributed since pulsar distances are not typically known with precisions better than the GW wavelength. For brevity, we omit the pulsar index $i$ in the derivations below.

In the single-source case, the likelihood for a single pulsar is
\begin{equation}
    q(d|z_{e}) = \frac{1}{2\pi\sigma^2_{n}} \int_0^{2\pi} \frac{d\psi}{2\pi}\ \exp\left\{-\frac{|d-z_{e}- z_{e}e^{i\psi}|^2}{2 \sigma^2_{n}}\right\}.
\end{equation}
This integral can be evaluated analytically, yielding the marginal distribution
\begin{equation}
q(d|z_{e}) = \frac{1}{2\pi\sigma^2_n} \exp\left\{-\frac{|d-z_{e}|^2 + |z_{e}|^2}{2\sigma^2_n}\right\} I_0\left[\frac{|z_{e}(d-z_{e})|}{\sigma^2_n}\right],
\label{eq:p_1s}
\end{equation}
where $I_0$ is the modified Bessel function of the first kind,
\begin{equation}
I_0(x) = \frac{1}{\pi}\int_0^\pi e^{x\cos\theta} d\theta.
\end{equation}

One might also want to consider an intermediate scenario in which one source dominates the signal, but the cumulative contribution from many fainter sources remains non-negligible and can be approximated as Gaussian. In this case, the signal can be decomposed as the sum of a dominant source and a stochastic background, so that $z = z_1 + z_{\rm stoc}$, and the Earth and pulsar terms become
\begin{equation}
    z_e = z_{e,1} + z_{e, {\rm stoc}} \quad \text{and} \quad z_p = z_{e,1}e^{i\psi} + z_{p,{\rm stoc}},
\end{equation}
where $z_{p,{\rm stoc}} \sim \mathcal{N}(0, \sigma_{p,{\rm stoc}})$. The Gaussian approximation to the likelihood remains the same as in Eq.~\ref{eq:p_1s}, since $2\sigma_p^2 = 2(\sigma^2_{p,1} + \sigma^2_{p,{\rm stoc}}) = \sum_s |z^s_{e}|^2$, while the full phase-marginalized likelihood becomes
\begin{equation}
\begin{split}
    q(d|z_{e}) =& \frac{1}{2\pi\sigma^2_N} \exp\left\{-\frac{|d-z_{e}|^2 + |z_{1,e}|^2}{2\sigma^2_N}\right\} \times \\ &I_0\left[\frac{|z_{1,e}(d-z_{e})|}{\sigma^2_N}\right],
    \label{eq:p1s_stoc}
\end{split}
\end{equation}
where we defined $\sigma^2_N \equiv \sigma^2_n + \sigma^2_{p, {\rm stoc}}$ to include all stochastic contributions. In this case, the relevant parameter controlling deviations from Gaussianity is $\varepsilon \equiv |z_{e,1}|^2/(2\sigma^2_N) = \sigma^2_{p,1}/\sigma^2_N$. Note that this $\varepsilon$ is not exactly the same as the one given in Eq.~\ref{eq:varepsilon}, since it is the ratio of the dominant source power as seen by the individual pulsar and the noise, and the definitions only coincide for a stochastic background. Nevertheless, we use $\varepsilon \equiv |z_{e,1}|^2/(2\sigma^2_N)$ in this appendix since there is no ambiguity given that we are only concerned with the likelihood in the point-source regime. In the example discussed here, we focus on the single-source case ($z=z_1$, $\sigma_N^2 = \sigma_n^2$), but the results can be easily generalized to the intermediate case by reinterpreting $\varepsilon$ as the ratio of the dominant source power to the total (GWB+noise) stochastic power.

Figure~\ref{fig:pdfs} shows the distribution $q$ for $\varepsilon = 0.5, 3$, with $\sigma^2_N(1+\varepsilon) = 1$ held fixed. The left panel corresponds to cross-sections along the real axis, while the right panel shows the two-dimensional distribution in the complex plane for $\varepsilon=5$. When $\varepsilon \lesssim 1$, the full distribution is well approximated by a Gaussian, while if $\varepsilon \gtrsim 1$, the distribution approaches a circle in the complex plane with radius $|z_e|$ centered on $z_e$, and significantly deviates from Gaussianity.

Since our primary goal is to measure the spherical harmonic coefficients $a_{\ell m}$ and $b_{\ell m}$, the single-pulsar likelihoods $q$ and $p$ are not the most relevant quantities to compare. Instead, we should examine the distributions of the estimators for the spherical harmonic coefficients. Even when the single-pulsar likelihood is highly non-Gaussian, the coefficient estimators combine data from $N_p$ pulsars and therefore approach Gaussian distributions as $N_p$ increases or $\varepsilon$ decreases. Lower multipoles $\ell$ effectively average over a larger number of pulsars and are consequently expected to be more Gaussian.

To illustrate this point, we will begin by considering the mean estimator $\hat{z} = \frac{1}{N_p}\sum_{i=1}^{N_p} d_i$ and take the Edgeworth expansion of the full distribution $q$. This estimator is unbiased under both distributions, since $\avg{\hat z}=z_e$. Also note that the real and imaginary components of the residual $d_i - z_e$ are identically distributed. Let $x = |z_e|\cos(\psi + \phi_e) + \Re(n)$ be the real part of the residual and $y$ the imaginary part, where $\psi$ is the unknown pulsar-term phase and $\phi_e$ is the (fixed) Earth-term phase. The variance of $x$ is given by
\begin{equation}
  \sigma^2_x = \frac{|z_{e,1}|^2}{2} + \sigma_N^2 \equiv \sigma_N^2\left(1+\varepsilon\right),
  \label{eq:sigma_x}
\end{equation}
where the factor of one half comes from the fact that averaging over a uniform distribution for $\psi$ gives $\avg{\cos^2\psi} = 1/2$. The variance of the mean estimator is $\sigma^2_{\hat{z}} = \sigma^2_N(1+\varepsilon)/N_p$ and the skewness vanishes due to the symmetry of $q$. The next term in the Edgeworth expansion is the fourth cumulant $\kappa_4$, which under the distribution $q$ is given by
\begin{equation}
  \kappa_4(x) \equiv \avg{x^4} - 3\avg{x^2}^2 = -\frac{3}{8} |z_{e,1}|^4.
  \label{eq:kappa_x}
\end{equation}
This results in a standardized excess kurtosis of $\gamma_2 (x)= \kappa_4(x)/\sigma_x^4 = -\frac{3}{2} \varepsilon^2/(1+\varepsilon)^2 = \gamma_2 (y)$. While $x$ and $y$ are uncorrelated, they are not independent due to the fact that the shared phase $\psi$ couples them through $\avg{\cos^2\psi \sin^2\psi} = 1/8$. The corresponding joint fourth-order cumulant is
\begin{equation}
\kappa_{22} \equiv \avg{x^2 y^2} - \avg{x^2}\avg{y^2} - 2\avg{xy}^2 = -\tfrac{1}{8}|z_{e,1}|^4,
\label{eq:kappa_22}
\end{equation}
where the last term vanishes since $\avg{xy}=0$ and the standardized value is $\kappa_{22}/\sigma_x^2\sigma_y^2= \gamma_2/3$. Cumulants are homogeneous of degree $n$, so $\kappa_n(\mathcal{W}_i x_i) = \mathcal{W}_i^n \kappa_n(x_i)$, and additive for independent variables, so
\begin{equation}
  \kappa_n\left(\sum_i \mathcal{W}_i x_i\right) = \sum_i \mathcal{W}_i^n \kappa_n(x_i).
  \label{eq:kappa_prop}
\end{equation}
For equal weights $\mathcal{W}_i = 1/N_p$ and iid variables $x_i$, we therefore conclude that the excess kurtosis is $\gamma_2(\hat x) = \gamma_2 (x)/N_p$ and the same holds for the cross-component $\kappa_{22}$.

To quantify how well the Gaussian approximates the full distribution, we compute the Kullback--Leibler (KL) divergence between the Edgeworth expansion of $q$ and the Gaussian approximation $p$ given in Eq.~\ref{eq:p_gauss}. The distribution $q$ can be approximated up to the fourth moment by
\begin{equation}
q(\hat{x}, \hat{y}) \approx \ p(\hat{x}) p(\hat{y}) \left[1 + \delta(\hat{x}, \hat{y})\right] ,
\end{equation}
where the small correction $\delta(\hat{x}, \hat{y})$ is given by
\begin{equation}
    \delta(\hat{x}, \hat{y}) = \frac{\gamma_2}{4!} He_4\left(\bar{x}\right) + \frac{\gamma_2}{4!} He_4\left(\bar{y}\right) + \frac{\gamma_2}{3 (2!)^2} He_2\left(\bar{x}\right)He_2\left(\bar{y}\right)
\end{equation}
and the standardized real and imaginary parts are defined as $\bar{x} \equiv \frac{\hat{x} - \avg{x}}{\sigma_{\hat{x}}}$ and similarly for $\bar{y}$.

The KL divergence between $q$ and $p$ is then given by
\begin{equation}
    \begin{split}
    D_{\rm KL} [q||p] \approx& \int d\hat{x} d\hat{y}\ p(\hat{x}) p(\hat{y}) \left[1+\delta\right] \ln \left[1+\delta\right] \\
    \approx& \int d\hat{x} d\hat{y}\ p(\hat{x}) p(\hat{y}) \left[\delta+ \frac{\delta^2}{2} \right] 
    \end{split}
\end{equation}
where we expanded the logarithm in the second approximation. To compute the value of the KL divergence, we first note that the integral over the first term $\delta$ vanishes due to the orthogonality of the Hermite polynomials. Similarly, the cross terms in the integral over the second term $\delta^2$ (involving the product of two Hermite polynomials) vanish due to orthogonality, and the remaining terms are only those $\propto He_4^2$ or $\propto He_2^2$. Evaluating this integral, leads to
\begin{equation}
    D_{\rm KL} [q||p] = \frac{\gamma_2^2}{18} = \frac{1}{8} \frac{\varepsilon^4}{\left(1+\varepsilon\right)^4}\cdot\frac{1}{N_p^2}
    \label{eq:KLleading}
\end{equation}
where we substituted the value of $\gamma_2$ for the mean estimator in the second equality. We therefore find that, at fixed $\varepsilon$, the KL divergence decays as $1/N_p^2$, while in the limit $\varepsilon\rightarrow \infty$, it saturates to the $N_p$-dependent constant $D_{\rm KL} \rightarrow \frac{1}{8 N_p^2}$. Since the correction to the Gaussian is required to be small, it is clear that the $\varepsilon \rightarrow \infty$ limit is beyond the range of validity of the Edgeworth expansion.

\begin{figure}[t]
    \centering
    \includegraphics[width=0.95\linewidth]{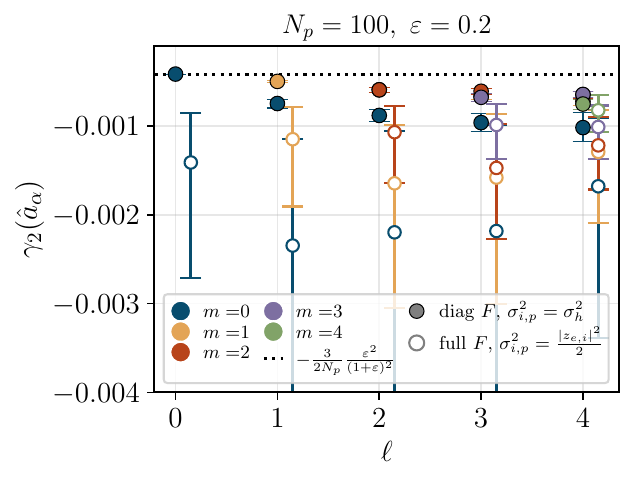}
    \caption{Excess kurtosis computed from Eq.~\ref{eq:g2} for a single source located in the $\hat{z}$ direction. Error bars indicate the range obtained by uniformly sampling pulsar positions on the sky.}
    \label{fig:kurt}
\end{figure}

We now extend this analysis to the spherical harmonic estimators, which we discuss in more detail in Sec.~\ref{sec:harm}. The maximum likelihood estimators are given by Eq.~\ref{eq:alm_MLE} are a weighted sum of the data $\hat{a}_{\alpha} = \sum_{i} \mathcal{W}_i^{\alpha} d_i$ with weights $\mathcal{W}^{\alpha}_i = \sum_{\beta} (F^{-1})_{\alpha \beta} Y^*_{\beta, i} w_i$, where $F_{\alpha\beta}$ is the Fisher matrix from Eq.~\ref{eq:fisher_aa}. We can in general compute the variance and excess kurtosis of the estimator using Eqs.~\ref{eq:sigma_x}, \ref{eq:kappa_x}, and \ref{eq:kappa_prop}, which gives
\begin{equation}
    \gamma_2(\hat{a}_\alpha) = -\frac{3}{8}\frac{\sum_i |z_{e,i}|^4\,|\mathcal{W}^{\alpha}_{i}|^4}{\left[(F^{-1})_{\alpha\alpha}\right]^2}.
    \label{eq:g2_general}
\end{equation}
In the limit of many uniformly distributed and equal $\varepsilon$ pulsars, the coupling matrix becomes diagonal and equal to Eq.~\ref{eq:fisher_aa}. In this case, the estimator simplifies to
\begin{equation}
    \hat{a}_{\alpha} \approx \frac{4\pi}{N_p} \sum_i d_i Y^*_{\alpha,i}.
    \label{eq:ahat_diag}
\end{equation}
Under this approximation, the standardized excess kurtosis is now
\begin{equation}
    \gamma_2 = -\frac{3}{2} \frac{\varepsilon^2}{\left(1 + \varepsilon \right)^2}\frac{\sum_i |Y_{\alpha,i}|^4}{\left[\sum_i |Y_{\alpha,i}|^2\right]^2}.
    \label{eq:g2}
\end{equation}
When all pulsars are identical and we estimate $a_{00}$, this expression reduces to the result for the mean estimator, as expected. The kurtosis is always negative (consistent with the shape shown in Fig.~\ref{fig:pdfs}) and attains a minimum value of $-3/2$ for $N_p=1$ in the limit of infinite SNR, although the Edgeworth expansion breaks down well before this regime is reached. 

Fig.~\ref{fig:kurt} shows the excess kurtosis for a face-on binary located in the $\hat{z}$ direction for $N_p=100$ pulsars with $\varepsilon=0.2$. We sample pulsar positions on the sky uniformly and compute the excess kurtosis for each configuration, resulting in the range shown by the error bars. The filled dots show the result in the diagonal approximation of Eq.~\ref{eq:g2} and equal $\varepsilon_i$ for all pulsars, while the unfilled circles show the result for the same configuration but including the full (non-diagonal) covariance matrix from Eq.~\ref{eq:fisher_aa}. The exact structure of the kurtosis as a function of $\ell$ and $m$ is a product of the assumed source sky location.

\begin{figure*}[h]
    \centering
    \includegraphics[width=0.65\linewidth]{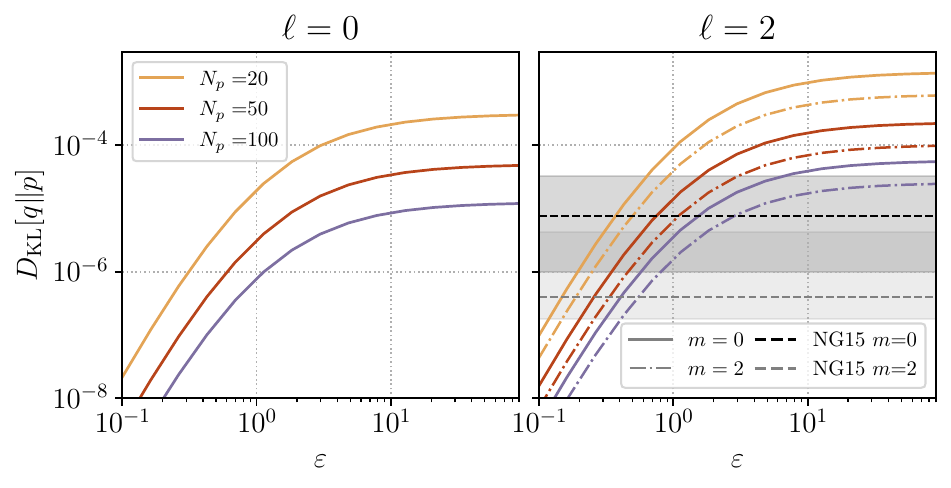}
    \caption{Leading-order KL divergence between the Gaussian approximation and the full distribution in the single-source scenario. Colored solid and dash-dotted lines show the KL divergence for $N_p = 20, 50, 100$ pulsars. Dashed horizontal lines and shaded regions indicate the result using the NANOGrav 15yr noise properties (median and 16th--84th percentile over source positions, phases, polarizations, and inclination angles). Left panel: $\ell=m=0$. Right panel: $\ell=2$ with $m=0$ (solid) and $m=2$ (dash-dotted).}
    \label{fig:DKL}
\end{figure*}

We then compute the KL divergence between the Gaussian approximation and the Edgeworth expansion of the full distribution. The colored solid and dash-dotted lines in Fig.~\ref{fig:DKL} show the KL divergence as a function of $\varepsilon$ for different numbers of pulsars in the diagonal approximation of Eq.~\ref{eq:g2}. We also include a more complete result using the noise properties of NANOGrav 15yr pulsars. In this case, we consider only the most sensitive frequency bin $f = 4$nHz and assume the entire GW signal is produced by a single source. We then sample the source sky position, phase, polarization, and inclination angles uniformly and compute the KL divergence for each realization, rescaling the signal power to keep the total $\rho^2_e$ constant. The median and 16th--84th percentile ranges over the monte carlo realizations are shown in dashed gray lines and shaded regions. We note that in this case, pulsars have unequal signal amplitudes $|z_{e,i}|^2$, which are accounted for in both the $|z_{e,i}|^2$ and $\mathcal{W}^{\alpha}_i$ terms in Eq.~\ref{eq:g2_general}.

We have shown in Eq.~\ref{eq:KLleading} that the KL divergence between the timing residual distribution for a single source marginalized over the pulsar phase and its Gaussian approximation is $\frac{1}{8} \frac{\varepsilon^4}{\left(1+\varepsilon\right)^4}$ for a single pulsar. Instead of computing the KL divergence for the distribution of the estimator $\hat{a}_\alpha$, we can also take an upper limit on the KL divergence for $N_p$ pulsars by considering the sum $\frac{1}{8} \frac{\varepsilon^4}{\left(1+\varepsilon\right)^4} N_p$, since each pulsar is independent. A standard metric in the field is to consider the Hellinger distance between two distributions~\cite{Lamb:2023jls, Laal:2024trp}, with $\lesssim 0.1$ typically considered a good approximation. Hence, we note that the KL divergence bounds the Hellinger distance from above, with $H^2[q||p] \leq \tfrac{1}{2} D_{\rm KL} [q||p]$~\cite{Weiss1991}. From the monte carlo realizations, we also compute the upper bound on the Hellinger distance, and find that the median upper limit across realizations is $H^2[q||p] \lesssim 0.05$ and $\lesssim 0.13$ for the $1\sigma$ upper bound. The corresponding KL divergence values for the $\hat{a}_\alpha$ estimator are necessarily lower, and we conclude that the Gaussian approximation is broadly applicable.

\end{document}